\documentclass[final,5p,times]{elsarticle}
\usepackage{graphicx,amsmath,mathtools}
\usepackage{hyperref}
\usepackage{bm,makecell}
\usepackage{siunitx}
\usepackage{color}
\usepackage{bold-extra}
\usepackage{multirow}
\usepackage[T1]{fontenc}
\usepackage{braket}
\usepackage{bm}
\usepackage{arydshln}
\usepackage{xcolor}
\usepackage{subcaption}
\usepackage{float}
\usepackage{mathtools}
\usepackage{soul}
\usepackage{fancyhdr}

\newcommand{\review}[1]{{\color{black}#1}}

\newtheorem{lemma}{Lemma}[section]
\DeclarePairedDelimiter{\ceil}{\lceil}{\rceil}

\newcommand\ddfrac[2]{\frac{\displaystyle #1}{\displaystyle #2}}

\newcommand\rhohat{\widehat{\rho}(\omega)}
\newcommand\rhohati{\widehat{\rho}(\Omega_i)}
\newcommand\nicolas[1]{{\color{black} #1}}

\RequirePackage[ruled]{algorithm2e}

\journal{Computer Physics Communications}

\fancypagestyle{specialfooter}{%
  \fancyhf{}
  
  \fancyfoot[L]{\small{\emph{Preprint submitted to Computer Physics Communications
} \\ LLNL-JRNL-866460}}
}

\begin{document}

\begin{frontmatter}

\title{Computing the QRPA Level Density with the Finite Amplitude Method}
\author[a]{A. Bjelčić\corref{author}}
\author[a]{N. Schunck}

\cortext[author] {Corresponding author.\\\textit{E-mail address:} bjelcic1@llnl.gov}
\address[a]{Nuclear and Chemical Science Division, Lawrence Livermore National Laboratory, Livermore, CA 94551, USA}

\date{}

\begin{abstract}
We describe a new algorithm to calculate the vibrational nuclear level density of 
an atomic nucleus. Fictitious perturbation operators that probe the response of 
the system are generated by drawing \nicolas{their} matrix elements from some probability 
distribution function. We use the Finite Amplitude Method to explicitly compute 
the response for each such sample. With the help of the Kernel Polynomial 
Method, we build an estimator of the vibrational level density and provide the 
upper bound of the relative error in the limit of infinitely many random samples. 
The new algorithm can give \nicolas{accurate} estimates of the vibrational level density. 
Since it is based on drawing multiple samples of perturbation operators, its 
computational implementation is naturally parallel and scales like the number 
of available processing units.
\end{abstract}


\end{frontmatter}

\thispagestyle{specialfooter}

\section{Introduction}


The first attempts to calculate the nuclear level density were based on the 
statistical model \cite{bethe1936attempt,bloch1954theory,ericson1960statistical,
bohr1998nucleara}. The formulas for the nuclear level density and spin distribution 
of excited states that are implemented in modern nuclear reaction codes such as 
TALYS \cite{koning2008talys1} or EMPIRE \cite{herman2007empire} are largely based on 
these early formulations. While such an approach gives the correct asymptotic, it 
relies on adjustable parameters and usually fails to capture the low-lying, discrete 
part of the spectrum. Direct methods based on the nuclear shell model are much 
more predictive but can only be applied in light-to-medium mass nuclei or with 
schematic interactions \cite{nakada1997total,horoi2003spin,alhassid2003nuclear,
nakada2008isospinprojected,senkov2010highperformance,senkov2011highperformance,
alhassid2015direct}.

What is often referred to as the combinatorial method is a good compromise between 
predictive power and feasibility \cite{hilaire2001combinatorial}. It is based on 
the general framework of the nuclear energy density functional theory, where 
several many-body methods can be leveraged to compute nuclear properties at a 
reasonable computational cost \cite{schunck2019energy}. The combinatorial method 
relies on counting nuclear excitations using generating function techniques
\cite{berger1974shell}. In the earlier version of the method, nuclear excited 
states were computed as $n$-particle $n$-holes excitations 
\cite{hilaire2001combinatorial}. The theory was extended to include vibrational 
and rotational states \cite{hilaire2006global,goriely2008improved} and the effect 
of nuclear excitation energy on the single-particle structure 
\cite{hilaire2012temperaturedependent}. Most recently, particle-hole excitations 
were replaced by the exact eigenvalues of the \nicolas{Quasiparticle Random Phase} (QRPA) matrix, giving a much more 
realistic description of low-lying states \cite{hilaire2023new}.

While the latest version of the combinatorial method contains a lot of new physics, 
it has become computationally expensive in deformed nuclei, where the QRPA matrix 
may easily reach sizes of $10^6 \times 10^6$ for realistic harmonic oscillator basis. 
Computing all these matrix elements is only doable on a \nicolas{case-by-case basis} or with 
additional approximations such as cutoffs in quasiparticle space or smaller bases 
\cite{terasaki2010selfconsistent,martini2016largescale}. The Finite Amplitude 
Method (FAM) offers a computationally efficient alternative to computing the QRPA
linear response for a given excitation operator without explicitly constructing
the QRPA matrix, but does not have direct access to the QRPA eigenfrequencies, i.e.
the vibrational spectrum
 \cite{nakatsukasa2007finite,avogadro2011finite}.

In this paper, we show that we can \nicolas{still} use the Finite Amplitude Method to efficiently 
probe the vibrational spectrum of nuclei of arbitrary size and compute estimates 
of the density of phonons with controlled precision.

In Section \ref{sec:qrpa}, we summarize very briefly the QRPA and linear response 
theory to introduce the vibrational level density. Our new solution method is outlined in Section
\ref{sec:method}. Practical aspects related to its 
implementation, which requires introducing an additional algorithm known as the 
Kernel Polynomial Method, are discussed in Section \ref{sec:practical}.
In Section \ref{sec:tests}, we illustrate the performance of the method with several benchmarks, both with 
synthetic or realistic data.


\section{QRPA level density}
\label{sec:qrpa}

In this section, we recall some of the basic properties of the QRPA matrix and of 
the linear response theory formalism which is used to compute the response of the 
nucleus to a given excitation operator.


\subsection{Notations}

Throughout the paper we use $A^\dagger$ for the Hermitian conjugate matrix, $A^T$ for the
transposed matrix and $A^*$ for the element-wise complex conjugate matrix of an arbitrary 
matrix $A\in\mathbb{C}^{m\times n}$. We also use MATLAB notation for diagonal 
matrices, i.e. if $\mathbf{x}\in\mathbb{C}^n$ is a vector, then 
$\operatorname{diag}[x]\in\mathbb{C}^{n\times n}$ is a diagonal matrix having 
elements of the vector $\mathbf{x}$ as diagonal elements, i.e. $\operatorname{diag}[x]_{ij} 
= \delta_{ij} x_i$, for $i,j=1,\dots,n$.
For a matrix 
$A\in\mathbb{C}^{m\times n}$, the Frobenius norm is denoted by: $||A||_F = 
\sqrt{\sum_{i=1}^m\sum_{j=1}^n |A_{ij}|^2}$, while the Euclidean norm of a 
vector $\mathbf{v}\in\mathbb{C}^{n}$ is $||\mathbf{v}||=\sqrt{\sum_{i=1}^n |v_i|^2}$. We denote the 
imaginary unit as $\mathfrak{i}$. For an $n$-dimensional random vector 
$\mathbf{X}$ that follows the multivariate Gaussian distribution with expected value 
$\boldsymbol{\mu}\in\mathbb{R}^n$ and covariance matrix 
$\Sigma\in\mathbb{R}^{n\times n}$, we use the notation: 
$\mathbf{X} \sim \mathcal{N}_n(\boldsymbol{\mu},\Sigma)$. For a random variable 
$X$, $\mathbb{E}[X]$ is its expected mean value and $\operatorname{Var}(X)$ its variance. 


\subsection{QRPA matrix}

\nicolas{The goal of this paper} is to compute the QRPA {level density function} 
$\rho(\omega)$, for all $\omega\geq 0$, defined as
\begin{equation}\label{Eq:LevelDensityDefinition}
    \rho(\omega) = \sum_{i=1}^{N_p} \delta(\omega-\Omega_i),
\end{equation}
where $N_p\in\mathbb{N}$ is the number of quasiparticle pairs $\mu<\nu$, and the 
$\Omega_i$ are the QRPA eigenfrequencies defined as the positive eigenvalues of the QRPA matrix $\mathcal{S}$ 
\begin{equation}\label{Eq:QRPAmatrix}
\mathcal{S} =
\left[\begin{matrix} +\boldsymbol{I} & \boldsymbol{0} \\ \boldsymbol{0} & -\boldsymbol{I} \end{matrix}\right]
\begin{bmatrix*}[c] A\phantom{^*} & B^{\phantom{*}} \\ B^* & A^* \end{bmatrix*},
\end{equation}
where the matrices $A,B\in \mathbb{C}^{N_p\times N_p}$ satisfy $A^\dagger = A$ and 
$B^T = B$ and involve the second derivatives of the energy functional with respect 
to the normal and pairing densities \cite{schunck2019energy,ring2004nuclear}. 
It is customary to introduce the QRPA eigenmodes $X,Y\in\mathbb{C}^{N_p\times N_p}$ 
to define the eigenvectors of the QRPA matrix. If 
$\Omega = \operatorname{diag}\left[\Omega_i\right]_{i=1}^{N_p} \in 
\mathbb{R}^{N_p\times N_p}$ is the diagonal matrix made of its eigenvalues and is 
such that $\Omega_i>0$ for each $i=1,\dots,N_p$, then one can show that
\begin{equation}\label{Eq:QRPAdiag1}
    \left[\begin{matrix} +\boldsymbol{I} & \boldsymbol{0} \\ \boldsymbol{0} & -\boldsymbol{I} \end{matrix}\right]
    \left[\begin{matrix} A\phantom{*} & B\phantom{*} \\ B^* & A^* \end{matrix}\right]
    =
    \left[\begin{matrix} X & Y^* \\ Y & X^* \end{matrix}\right]
    \left[\begin{matrix} +\Omega & \boldsymbol{0} \\ \boldsymbol{0} & -\Omega \end{matrix}\right]
    \left[\begin{matrix} X & Y^* \\ Y & X^* \end{matrix}\right]^{-1},
\end{equation}
with
\begin{equation}\label{Eq:QRPAdiag2}
    \left[\begin{matrix} X & Y^* \\ Y & X^* \end{matrix}\right]^{-1}
    =
    \left[\begin{matrix} +\boldsymbol{I} & \boldsymbol{0} \\ \boldsymbol{0} & -\boldsymbol{I} \end{matrix}\right]
    \left[\begin{matrix} X & Y^* \\ Y & X^* \end{matrix}\right]^{\dagger}
    \left[\begin{matrix} +\boldsymbol{I} & \boldsymbol{0} \\ \boldsymbol{0} & -\boldsymbol{I} \end{matrix}\right].
\end{equation}

In the context of numerical linear algebra, the function $\rho(\omega)$ is known 
as the spectral density of the QRPA matrix $\mathcal{S}$. The practical 
calculation of this function in the case of the QRPA matrix \eqref{Eq:QRPAmatrix} 
faces two problems
\begin{itemize}
\item In spherical nuclei, the QRPA matrix $\mathcal{S}$ can be 
easily constructed and diagonalized as in Eqs. \eqref{Eq:QRPAdiag1} and 
\eqref{Eq:QRPAdiag2}. In that case, one has an explicit access to the QRPA 
eigenfrequencies $\{\Omega_i\}_{i=1}^{N_p}$, and therefore the computation of the 
level density function $\rho(\omega)$ becomes trivial. However, in heavy deformed 
nuclei the dimension $N_p$ becomes prohibitively large as it can reach $N_p\sim 10^6$ 
for large harmonic oscillator bases. In those cases, the direct determination and 
diagonalization of the QRPA matrix has only been achieved in a few cases 
\cite{terasaki2010selfconsistent,martini2016largescale}.
\item Although numerous numerical methods for estimating the spectral density exist 
\cite{dinapoli2016efficient,lin2016approximating}, they all assume that the matrix 
for which the spectral density is computed is Hermitian. In our case, the QRPA 
matrix $\mathcal{S}$ is not Hermitian and the matrix of its eigenvectors is not 
unitary but involves the orthonormality condition of Eq. \eqref{Eq:QRPAdiag2}.
Therefore, those spectral density methods cannot be directly used.
\end{itemize}


\subsection{Linear response theory}

While the QRPA approximation is the small amplitude limit of the time-dependent 
Hartree-Fock-Bogoliubov theory, linear response theory is the same limit in the 
presence of an external, time-dependent field $\hat{F}(t)$ \cite{ring2004nuclear}. 
In the quasiparticle basis \review{represented by the quasiparticle creation and 
annihilation operators $\hat{\beta}_{\mu}^{\dagger}$ and $\hat{\beta}_{\nu}$}, 
this field is represented as \cite{schunck2019energy}
\begin{equation}\label{Eq:Foperator}
    \review{
    \hat{F} = \frac{1}{2}\sum_{\mu\nu} \left[ 
     F^{11}_{\mu\nu}  \hat{\beta}_\mu^\dagger \hat{\beta}_\nu +
     F^{20}_{\mu\nu}  \hat{\beta}_\mu^\dagger \hat{\beta}_\nu^\dagger +
     F^{02}_{\mu\nu}  \hat{\beta}_\mu \hat{\beta}_\nu -
     F^{11*}_{\mu\nu} \hat{\beta}_\mu \hat{\beta}_\nu^\dagger
     \right].
     }
\end{equation}
Let $F^{20},F^{02}\in\mathbb{C}^{N_p}$ be the vectorized
\review{strict upper triangle parts ($\mu<\nu$) of the}
matrices 
$F^{20}_{\mu\nu},F^{02}_{\mu\nu}$. For $\omega\in\mathbb{R}$ and smearing 
$\gamma>0$, we denote the complex frequency 
$\omega_\gamma = \omega + \gamma\mathfrak{i}$ in the upper complex plane.

\review{In the small amplitude limit, the external perturbation induces
oscillations of the Hamiltonian $\delta H^{20}_{\mu\nu}(\omega_\gamma)$ and $ \delta H^{02}_{\mu\nu}(\omega_\gamma)$
according to the equations \cite{avogadro2011finite},
\begin{align}\label{Eq:QFAMequations}
    (E_\mu+E_\nu-\omega_\gamma) X_{\mu\nu}(\omega_\gamma) + \delta H^{20}_{\mu\nu}(\omega_\gamma) &= F^{20}_{\mu\nu}(\omega_\gamma),
    \nonumber \\
    (E_\mu+E_\nu+\omega_\gamma) Y_{\mu\nu}(\omega_\gamma) + \delta H^{02}_{\mu\nu}(\omega_\gamma) &= F^{02}_{\mu\nu}(\omega_\gamma),
\end{align}
where the antisymmetric matrices $X_{\mu\nu}(\omega)$ and $Y_{\mu\nu}(\omega)$ are the corresponding forward and backward amplitudes.
Let the
vectors $X(\omega_\gamma)$, $Y(\omega_\gamma)\in\mathbb{C}^{N_p}$ denote the
vectorized strict upper triangle parts of the matrices
$X_{\mu\nu}(\omega_\gamma),Y_{\mu\nu}(\omega_\gamma)$.
By expanding $\delta H^{20}_{\mu\nu}(\omega_\gamma)$ and $\delta H^{02}_{\mu\nu}(\omega_\gamma)$ in terms of the amplitudes $X_{\mu\nu}(\omega_\gamma)$ and $Y_{\mu\nu}(\omega_\gamma)$
we obtain the linear response equation,
\begin{equation}\label{Eq:LinearResponseEquation}
    \left(
    \begin{bmatrix} A\phantom{*} & B\phantom{*} \\ B^* & A^* \end{bmatrix}
    - \omega_\gamma
    \begin{bmatrix} +\boldsymbol{I} & \boldsymbol{0} \\ \boldsymbol{0} & -\boldsymbol{I} \end{bmatrix}
    \right)
    \begin{bmatrix} X(\omega_\gamma) \\ Y(\omega_\gamma) \end{bmatrix} =
    - \begin{bmatrix} F^{20} \\ F^{02} \end{bmatrix}.
\end{equation}

Traditionally,
Eq.~\eqref{Eq:LinearResponseEquation} is solved by
explicitly constructing and diagonalizing the matrix $\mathcal{S}$ of Eq.~\eqref{Eq:QRPAmatrix}, which can become computationally prohibitive for heavy deformed nuclei.
However, it is in fact possible to solve for the 
$X(\omega_\gamma)$ and $Y(\omega_\gamma)$ without having to compute the matrix elements of $\mathcal{S}$ using the Finite Amplitude Method (FAM) \cite{nakatsukasa2007finite,
avogadro2011finite}.
The key idea of the FAM is to iteratively solve 
Eq.~\eqref{Eq:QFAMequations}, independently for
each frequency $\omega$ on a chosen mesh.
As was shown in Ref.~\cite{bjelcic2022chebyshev},
the FAM actually only accesses the mapping
\begin{equation}\label{Eq:QRPAmapping}
    \begin{bmatrix} x \\ y \end{bmatrix} \in \mathbb{C}^{2N_p}
    \mapsto
    \begin{bmatrix} A\phantom{*} & B\phantom{*} \\ B^* & A^* \end{bmatrix}
    \begin{bmatrix} x \\ y \end{bmatrix} \in \mathbb{C}^{2N_p},
\end{equation}
and therefore the
FAM could be classified as a matrix-free method for solving the linear system \eqref{Eq:LinearResponseEquation}.
In the rest of this paper, we will 
make extensive use of this property and design a method to estimate the spectral density 
$\rho(\omega)$ by only using the mapping \eqref{Eq:QRPAmapping}.
}

For any given operator $\hat{F}$, we define the strength function 
$S(\omega_\gamma,\hat{F})$ as
\begin{equation}
    S(\omega_\gamma,\hat{F}) = 
    \begin{bmatrix} F^{20} \\ F^{02} \end{bmatrix}^\dagger 
    \begin{bmatrix} X(\omega_\gamma) \\ Y(\omega_\gamma) \end{bmatrix},
\end{equation}
and the response function ${dB(\omega,\hat{F})}/{d\omega}$ as
\begin{equation}
    \frac{dB(\omega,\hat{F})}{d\omega} = 
    \lim_{\gamma\to 0^+} \frac{-1}{\pi} \operatorname{Im}\left[ S(\omega_\gamma,\hat{F}) \right].
\end{equation}
One can show that the response function can be written as
\begin{equation}\label{Eq:ResponseFunction}
    \frac{dB(\omega,\hat{F})}{d\omega} = \sum_{i=1}^{N_p} |\langle i | \hat{F} | 0 \rangle |^2 \delta(\omega-\Omega_i)
    -
    \sum_{i=1}^{N_p} |\langle 0 | \hat{F} | i \rangle |^2 \delta(\omega+\Omega_i),
\end{equation}
where $|0\rangle$ is the QRPA ground state, and 
$|i\rangle = \hat{O}_i^\dagger |0\rangle $ \nicolas{is the $i$-th excited state defined from the} phonon creation operators 
$\hat{O}_i^\dagger$ 
\begin{equation}\label{Eq:PhononOperator}
    \hat{O}_i^\dagger = \sum_{\mu<\nu} \left[ 
     X^{i}_{\mu\nu} \hat{\beta}_\mu^\dagger \hat{\beta}_\nu^\dagger -
     Y^{i}_{\mu\nu} \hat{\beta}_\nu \hat{\beta}_\mu
     \right].
\end{equation}
Using Eqs.~\eqref{Eq:Foperator} and \eqref{Eq:PhononOperator} we can compute 
the transition strengths \cite{hinohara2013lowenergy}
\begin{align}
    T_i(\hat{F}) = \langle i | \hat{F} | 0 \rangle = \langle \Phi_0 | [\hat{O}_i,\hat{F}] | \Phi_0 \rangle & =
    \sum_{\mu<\nu} \left[
    \big( X_{\mu\nu}^{i} \big)^* F_{\mu\nu}^{20}
    +
    \big( Y_{\mu\nu}^{i} \big)^* F_{\mu\nu}^{02} \right],
    \nonumber
\\
    \widetilde{T}_i(\hat{F}) = \langle 0 | \hat{F} | i \rangle = \langle \Phi_0 | [\hat{O}_i^\dagger,\hat{F}] | \Phi_0 \rangle & =
    \sum_{\mu<\nu} \left[
    \phantom{\big(}Y_{\mu\nu}^{i}\phantom{\big)^*} F_{\mu\nu}^{20}
    +
    \phantom{\big(}X_{\mu\nu}^{i}\phantom{\big)^*} F_{\mu\nu}^{02} \right],
\end{align}
where $|\Phi_0\rangle$ is the HFB state. \nicolas{A shorthand, vectorized form of these transition 
strengths reads}
\begin{equation}\label{Eq:iF0_0Fi}
    \begin{bmatrix*}[l] T(\hat{F}) \\ \widetilde{T}(\hat{F}) \end{bmatrix*} =
    \begin{bmatrix} X & Y^* \\ Y & X^* \end{bmatrix}^\dagger
    \begin{bmatrix} F^{20} \\ F^{02} \end{bmatrix}.
\end{equation}

Using Eq.~\eqref{Eq:QRPAdiag2}, we can also invert the previous equation:
\begin{equation}
    \begin{bmatrix} F^{20} \\ F^{02} \end{bmatrix} =
    \begin{bmatrix} +\boldsymbol{I} & \boldsymbol{0} \\ \boldsymbol{0} & -\boldsymbol{I} \end{bmatrix}
    \begin{bmatrix} X & Y^* \\ Y & X^* \end{bmatrix}
    \begin{bmatrix} +\boldsymbol{I} & \boldsymbol{0} \\ \boldsymbol{0} & -\boldsymbol{I} \end{bmatrix}
    \begin{bmatrix*}[l] T(\hat{F}) \\ \widetilde{T}(\hat{F}) \end{bmatrix*}.
\end{equation}
Notice that if we choose the excitation operator $\hat{F'}$ (i.e. 
picked $F'^{20}_{\mu,\nu},F'^{02}_{\mu,\nu}\in\mathbb{C}$) such that
\begin{equation}\label{Eq:Fhypothetical}
     \begin{bmatrix} +F'^{20} \\ -F'^{02} \end{bmatrix} =
    \begin{bmatrix} X & Y^* \\ Y & X^* \end{bmatrix}
    \begin{bmatrix} \textbf{1} \\ \textbf{1} \end{bmatrix},
\end{equation}
where $\textbf{1}=[1,\dots,1]^T \in \mathbb{R}^{N_p}$, then we would have:
\begin{equation}
    |T_i(\hat{F}')|^2 = |\langle i | \hat{F'} | 0 \rangle|^2 = 1 \quad \text{and} \quad |\widetilde{T}_i(\hat{F}')|^2 =|\langle 0  | \hat{F}' | i \rangle|^2 = 1,
\end{equation} 
for all $i=1,\dots,N_p$, and consequently the response function for that specific 
operator $\hat{F'}$ would be equal to the QRPA level density function
\begin{equation}
    \frac{dB(\omega,\hat{F'})}{d\omega} = \rho(\omega), \qquad \forall\omega\geq 0.
\end{equation}
In practice, such operator $\hat{F'}$ is impossible to construct because it 
essentially involves computing the sum of all eigenvectors in 
Eq.~\eqref{Eq:Fhypothetical}, and therefore requires the full diagonalization of 
the QRPA matrix. However, this simple example demonstrates that by carefully 
picking the excitation operator and afterwards solving only the linear response 
equation \eqref{Eq:LinearResponseEquation}, one can, at least in principle, obtain 
an approximation of the QRPA level density function. Therefore, the main question 
is how to efficiently construct an operator $\hat{F'}$ which excites all the 
eigenmodes with equal strength.


\section{Method description}
\label{sec:method}

We consider a collection of $N_s$ random excitation operators $\hat{F}_{i_s}$, for 
$1\leq i_s \leq N_s$, with matrix elements $F^{20}_{i_s},F^{02}_{i_s}$ drawn from a multivariate Gaussian distribution. We will show how to estimate the QRPA level density function $\rho(\omega)$ by introducing 
a specific averaging procedure over the collection of the obtained response functions 
${dB(\omega,\hat{F}_{i_s})}/{d\omega}$ computed for each sample operator. Before we proceed, 
we recall a well-known Lemma describing how a Gaussian random vector is affected by 
a linear transformation.

\begin{lemma}\label{Lemma:LinearTransformationOfMultivariateGaussian}
    Let $\mathbf{X}\sim \mathcal{N}_n(\boldsymbol{0},\Sigma)$ be a $n$-dimensional 
    random vector having multivariate Gaussian distribution with zero expected value 
    and covariance matrix $\Sigma$. Then for any matrix $A\in\mathbb{R}^{m\times n}$, 
    the $m$-dimensional random vector $A\mathbf{X}$ has a multivariate Gaussian 
    distribution with zero expected value and covariance matrix $A\Sigma A^T$, i.e. 
    $A\mathbf{X}\sim \mathcal{N}_m(\boldsymbol{0},A\Sigma A^T)$.
\end{lemma}


\subsection{Randomized excitation operator}

According to Theorem 1. in Appendix B from Ref. \cite{bjelcic2022chebyshev}, there 
exist $C,D\in\mathbb{C}^{N_p\times N_p}$ unitary matrices and diagonal matrix $\theta =\operatorname{diag}[\theta_i]_{i=1}^{N_p}$, for $\theta_1,\dots,\theta_{N_p}\geq 0$, 
such that
\begin{equation}
    X = D \cosh{\theta} C
    \quad \text{and} \quad
    Y = D^* \sinh{\theta} C.
\end{equation}
Inserting the previous relations into Eq.~\eqref{Eq:iF0_0Fi}, we obtain:
\begin{align}\label{Eq:iF0_0Fi_CD}
    \begin{bmatrix*}[l] T(\hat{F}) \\ \widetilde{T}(\hat{F}) \end{bmatrix*}
   & =
    \begin{bmatrix} X & Y^* \\ Y & X^* \end{bmatrix}^\dagger
    \begin{bmatrix} F^{20} \\ F^{02} \end{bmatrix} \nonumber
\\
   & =
    \begin{bmatrix}
        C^\dagger & \boldsymbol{0} \\
        \boldsymbol{0} & C^T
    \end{bmatrix}
    \begin{bmatrix}
        \cosh\theta & \sinh\theta \\
        \sinh\theta & \cosh\theta
    \end{bmatrix}
    \begin{bmatrix}
        D^\dagger & \boldsymbol{0} \\
        \boldsymbol{0} & D^T
    \end{bmatrix}
    \begin{bmatrix} F^{20} \\ F^{02} \end{bmatrix}.
\end{align}
Suppose that $F^{20},F^{02}\in\mathbb{C}^{N_p}$ are random complex vectors, such 
that \nicolas{all} the components
\begin{equation}\label{Eq:GaussianComponents}
\operatorname{Re}[F^{20}_i], \text{ } 
\operatorname{Re}[F^{02}_i], \text{ } 
\operatorname{Im}[F^{20}_i], \text{ } 
\operatorname{Im}[F^{02}_i] \sim \mathcal{N}_1(0,\sigma^2)    ,
\end{equation}
are independent and identically distributed Gaussian random variables with zero expected 
value and \nicolas{the same} variance $\sigma^2>0$, for each $i=1,\dots,N_p$. Then it is easy to show 
that the $4N_p$-dimensional random vector
\begin{equation}\label{Eq:F20F02randomVariable}
    \begin{bmatrix}
    \operatorname{Re}F^{20} \\
    \operatorname{Re}F^{02} \\
    \operatorname{Im}F^{20} \\
    \operatorname{Im}F^{02} \\
    \end{bmatrix}
    \sim \mathcal{N}_{4N_p}(\boldsymbol{0},\sigma^2 \boldsymbol{I}),
\end{equation}
has multivariate Gaussian distribution with zero expected value and covariance 
matrix $\sigma^2 \boldsymbol{I}$.

Let us introduce the real and imaginary parts: $C_1,C_2,D_1,D_2 \in 
\mathbb{R}^{N_p\times N_p}$ of $C$ and $D$ matrices, i.e. $C=C_1+C_2\mathfrak{i}$ and $D=D_1+D_2\mathfrak{i}$. It is easy to see that 
Eq.~\eqref{Eq:iF0_0Fi_CD} can be recast into
\begin{equation}
    \begin{bmatrix*}[l]
    \operatorname{Re}T(\hat{F}) \\
    \operatorname{Re}\widetilde{T}(\hat{F})\\
    \operatorname{Im}T(\hat{F}) \\
    \operatorname{Im}\widetilde{T}(\hat{F}) \\
    \end{bmatrix*}
    =
    \boldsymbol{\mathcal{{X}}}
    \begin{bmatrix}
    \operatorname{Re}F^{20} \\
    \operatorname{Re}F^{02} \\
    \operatorname{Im}F^{20} \\
    \operatorname{Im}F^{02} \\
    \end{bmatrix},
\end{equation}
where the real matrix $\boldsymbol{\mathcal{{X}}}\in\mathbb{R}^{4N_p\times 4N_p}$
is given by
\begin{equation}
\boldsymbol{\mathcal{{X}}} = 
\boldsymbol{\mathcal{C}}
    \begin{bmatrix}
        \cosh\theta & \sinh\theta & \boldsymbol{0} & \boldsymbol{0} \\
        \sinh\theta & \cosh\theta & \boldsymbol{0} & \boldsymbol{0} \\
        \boldsymbol{0} & \boldsymbol{0} & \cosh\theta & \sinh\theta  \\
        \boldsymbol{0} & \boldsymbol{0} & \sinh\theta & \cosh\theta  
    \end{bmatrix}
\boldsymbol{\mathcal{D}},
\end{equation}
with 
\begin{align}
\boldsymbol{\mathcal{C}}
& = \begin{bmatrix}
        C_1^T & \boldsymbol{0} & C_2^T & \boldsymbol{0} \\
        \boldsymbol{0} & C_1^T & \boldsymbol{0} & -C_2^T \\
        -C_2^T & \boldsymbol{0} & C_1^T & \boldsymbol{0} \\
        \boldsymbol{0} & C_2^T & \boldsymbol{0} & C_1^T 
    \end{bmatrix} \\
\boldsymbol{\mathcal{D}}
& =  \begin{bmatrix}
        D_1^T & \boldsymbol{0} & D_2^T & \boldsymbol{0} \\
        \boldsymbol{0} & D_1^T & \boldsymbol{0} & -D_2^T \\
        -D_2^T & \boldsymbol{0} & D_1^T & \boldsymbol{0} \\
        \boldsymbol{0} & D_2^T & \boldsymbol{0} & D_1^T 
    \end{bmatrix}.
\end{align}
Both matrices $\boldsymbol{\mathcal{C}}$ and $\boldsymbol{\mathcal{D}}$ are 
orthogonal since $C$ and $D$ are unitary. 

According to Lemma \ref{Lemma:LinearTransformationOfMultivariateGaussian}, we know 
that the following $4N_p$-dimensional random vector,
\begin{equation}\label{Eq:RandomVectoriF0_0Fi}
    \begin{bmatrix*}[l]
    \operatorname{Re}T(\hat{F}) \\
    \operatorname{Re}\widetilde{T}(\hat{F})\\
    \operatorname{Im}T(\hat{F}) \\
    \operatorname{Im}\widetilde{T}(\hat{F}) \\
    \end{bmatrix*}
    \sim
    \mathcal{N}_{4N_p}(\boldsymbol{0},{\Sigma}),
\end{equation}
has multivariate Gaussian distribution with zero expected value and (real) covariance matrix 
$\Sigma = \boldsymbol{\mathcal{X}} (\sigma^2 \boldsymbol{I}) \boldsymbol{\mathcal{X}}^T$.
We can easily extract an arbitrary component of the random vector 
\eqref{Eq:RandomVectoriF0_0Fi}. As an example, for any $j=1,\dots,N_p$, we have
\begin{equation}
    \operatorname{Re} T_j(\hat{F}) \equiv 
    \operatorname{Re}[ \langle j | \hat{F} | 0 \rangle ] = 
    \begin{bmatrix*}[l]
        \boldsymbol{e}_j \\ \boldsymbol{0} \\ \boldsymbol{0} \\ \boldsymbol{0}
    \end{bmatrix*}^T
    \begin{bmatrix*}[l]
    \operatorname{Re}T(\hat{F}) \\
    \operatorname{Re}\widetilde{T}(\hat{F})\\
    \operatorname{Im}T(\hat{F}) \\
    \operatorname{Im}\widetilde{T}(\hat{F}) \\
    \end{bmatrix*}.
\end{equation}
Invoking Lemma \ref{Lemma:LinearTransformationOfMultivariateGaussian} again, we see 
that the random variable $\operatorname{Re}[ \langle j | \hat{F} | 0 \rangle ] \sim 
\mathcal{N}(0,\sigma_j^2)$ has Gaussian distribution with zero expected value and 
variance $\sigma_j^2$ given by
\begin{equation}
\sigma_j^2 = 
    \begin{bmatrix*}[l]
        \boldsymbol{e}_j \\ \boldsymbol{0} \\ \boldsymbol{0} \\ \boldsymbol{0}
    \end{bmatrix*}^T
    \Sigma\ 
    \begin{bmatrix*}[l]
        \boldsymbol{e}_j \\ \boldsymbol{0} \\ \boldsymbol{0} \\ \boldsymbol{0}
    \end{bmatrix*}.
\end{equation}
Using the result $\Sigma = \boldsymbol{\mathcal{X}} (\sigma^2 \boldsymbol{I}) 
\boldsymbol{\mathcal{X}}^T$, this can be easily be computed explicitly as
\begin{equation}\label{eq:sigma_i}
\sigma_{j}^{2} = \sigma^2 + 2\sigma^2 || Y\boldsymbol{e}_j ||^2.
\end{equation}
Repeating 
a similar procedure for the other components of the random vector 
\eqref{Eq:RandomVectoriF0_0Fi}, we find that for any $i=1,\dots,N_p$, 
\begin{equation}\label{Eq:TransitionMatrixElementsDistributions}
    \operatorname{Re}[ \langle i | \hat{F} | 0 \rangle ], \text{ }
    \operatorname{Re}[ \langle 0 | \hat{F} | i \rangle ], \text{ }
    \operatorname{Im}[ \langle i | \hat{F} | 0 \rangle ], \text{ }
    \operatorname{Im}[ \langle 0 | \hat{F} | i \rangle ]
    \sim \mathcal{N}_1(0,\sigma_i^2)
\end{equation}
with the variance $\sigma_i^2$ given by Eq.~\eqref{eq:sigma_i}. It is worth mentioning 
that all these components are not necessarily independent random variables. 

We can now for any $i=1,\dots,N_p$ compute the following expected value
\begin{equation}
    \mathbb{E}\Big[|\langle i | \hat{F} | 0 \rangle|^2\Big] =
    \mathbb{E}\Big[ \operatorname{Re}[\langle i | \hat{F} | 0 \rangle]^2 \Big] 
    +
    \mathbb{E}\Big[ \operatorname{Im}[\langle i | \hat{F} | 0 \rangle]^2 \Big] .
\end{equation}
Since for any random variable $X$ with zero expected value
its variance can be expressed as  $\mathrm{Var}(X)=\mathbb{E}[X^2]$, result \eqref{Eq:TransitionMatrixElementsDistributions} yields:
\begin{equation}
    \mathbb{E}\Big[|\langle i | \hat{F} | 0 \rangle|^2\Big] 
    = \sigma_i^2 + \sigma_i^2 
    \nonumber \\
    = 2\sigma^2 + 4\sigma^2 ||Y\boldsymbol{e}_i||^2.
\end{equation}
The same procedure can be repeated for $|\langle 0 | \hat{F} | i \rangle|^2$. Overall, we conclude that if one draws the elements of $F^{20}$ and $F^{02}$ from a Gaussian distribution as in Eq.~\eqref{Eq:GaussianComponents}, then for any $i=1,\dots,N_p$ there holds:
\begin{equation}\label{Eq:FinalExpectedValueResult}
\mathbb{E}\Big[|\langle i | \hat{F} | 0 \rangle|^2\Big]
=
\mathbb{E}\Big[|\langle 0 | \hat{F} | i \rangle|^2\Big]
= 4\sigma^2 \left( \frac{1}{2} +
 ||Y\boldsymbol{e}_i||^2 \right).
\end{equation}


\subsection{Level density estimator}

Recall that the response function expression \eqref{Eq:ResponseFunction} for 
non-negative frequencies $\omega\geq 0$ reads
\begin{equation}\label{Eq:RandomVariable1}
    \frac{dB(\omega\geq 0,\hat{F})}{d\omega} = \sum_{i=1}^{N_p} |\langle i | \hat{F} | 0 \rangle|^2 \delta(\omega-\Omega_i).
\end{equation}
Notice that now the response function is a random variable because the excitation 
operator $\hat{F}$ is a random variable. Also notice that the following integral 
over non-negative frequencies
\begin{equation}\label{Eq:RandomVariable2}
    \int_0^{+\infty} \frac{dB(\omega'\geq 0,\hat{F})}{d\omega'} d\omega'
    =
    \sum_{i=1}^{N_p} |\langle i | \hat{F} | 0 \rangle|^2,
\end{equation}
is also a random variable.

Since both random variables in Eqs.~\eqref{Eq:RandomVariable1} and 
\eqref{Eq:RandomVariable2} are well defined, we can compute their expected values. 
Therefore, we can define the QRPA {level density estimator} $\rhohat$, for 
non-negative frequencies $\omega\geq 0$, as
\begin{equation}\label{Eq:LevelDensityEstimatorDefinition}
    \rhohat =
    N_p
    \ddfrac
    {\mathbb{E}\left[ \phantom{\int_{0}^{+\infty}} \frac{dB(\omega\phantom{'}\geq 0,\hat{F})}{d\omega} \phantom{.} \phantom{d\omega'} \right]}
    {\mathbb{E}\left[          \int_{0}^{+\infty}  \frac{dB(\omega'          \geq 0,\hat{F})}{d\omega'} \phantom{.}          d\omega'  \right]}.
\end{equation}
Inserting Eq.~\eqref{Eq:FinalExpectedValueResult} into Eq.~\eqref{Eq:RandomVariable1} 
and Eq.~\eqref{Eq:RandomVariable2}, we get
\begin{equation}
    \mathbb{E}\left[
    \frac{dB(\omega\geq 0,\hat{F})}{d\omega}
    \right]
    = 4\sigma^2 \sum_{i=1}^{N_p} \left( \frac{1}{2} + ||Y\boldsymbol{e}_i||^2 \right) \delta(\omega-\Omega_i).
\end{equation}
and
\begin{equation}
    \mathbb{E}\left[ \int_{0}^{+\infty} 
    \frac{dB(\omega'\geq 0,\hat{F})}{d\omega'}d\omega'
    \right]
    = 4\sigma^2 N_p \left( \frac{1}{2} + \frac{||Y||_F^2}{N_p} \right)
    .
\end{equation}

Now we can express the estimator $\rhohat$ as
\begin{equation}\label{Eq:RhoEstimatorExpression}
    \rhohat = \sum_{i=1}^{N_p}
\left(\frac{ \frac{1}{2} + {||Y\boldsymbol{e}_i||^2} }{ \frac{1}{2} + \frac{||Y||_F^2}{N_p} }\right)
    \delta(\omega-\Omega_i).
\end{equation}
In the interpretation where the Dirac delta function is a non-negative function of infinitesimal
width and unit area, we can easily show that the absolute error $| \rhohat - \rho(\omega) |$ with respect 
to the true QRPA level density \eqref{Eq:LevelDensityDefinition} is bounded as
\begin{equation}\label{Eq:ErrorEstimator}
    | \rhohat - \rho(\omega) |
    \leq
    \rho(\omega)
    \max_{i=1,\dots,N_p}\varepsilon_i
    ,
\end{equation}
with 
\begin{equation}\label{Eq:epsilon_i}
    \varepsilon_i = \left|\frac{ {||Y\boldsymbol{e}_i||^2} - \frac{||Y||_F^2}{N_p} }{ \frac{1}{2} + \frac{||Y||_F^2}{N_p} }\right|.
\end{equation}
The relative error bound for any $\omega\geq 0$ is then
\begin{equation}\label{Eq:epsiDefinition}
   \left| \frac{ \rhohat - \rho(\omega) }{ \rho(\omega) } \right|
   \leq \max_{i=1,\dots,N_p}  \varepsilon_i .
\end{equation}
Assuming for simplicity that the eigenfrequencies $\Omega_i$ are non degenerate, we 
see that as one approaches the given QRPA eigenfrequency  $\omega\rightarrow\Omega_i$, 
the local relative error is
\begin{equation}\label{Eq:relErrorAtOmegai}
\lim_{\omega\rightarrow\Omega_i} 
\left| \frac{ \rhohat - \rho(\omega) }{ \rho(\omega) } \right|= \varepsilon_i ,
    \quad\forall i=1,\dots,N_p  .
\end{equation}
It is worth emphasizing that \eqref{Eq:epsiDefinition} and \eqref{Eq:relErrorAtOmegai} hold only
approximately if the Dirac delta function is
approximated by a finite width function.

If we imagine the level density function $\rho(\omega)$ as the linear response 
function with all transition probabilities being equal to one, we can see that the 
level density estimator $\rhohat$ is the linear response function 
with the proper QRPA eigenfrequencies $\Omega_i$ but transition probabilities with 
relative error bounded by $\varepsilon_i$ for each $\Omega_i$. We also notice that 
$\varepsilon_i$ is a measure of how much the squared norm of the $i$-th column of the 
$Y$ matrix deviates from the average value over all columns. In practice, the norm 
$||Y||_F$ is relatively small because the matrix $Y$ describes the ground-state 
correlations \cite{ring2004nuclear}.
If we expand the residual interaction $\delta H^{20}_{\mu\nu}$ and $\delta H^{02}_{\mu\nu}$
in terms amplitudes $X_{\mu\nu}$ and $Y_{\mu\nu}$:
\begin{align}
    \delta H^{20}_{\mu\nu} & = -(E_\mu+E_\nu)X_{\mu\nu}
    + \sum_{\mu'<\nu'} \left[ A_{\mu\nu,\mu'\nu'} X_{\mu'\nu'} +  B_{\mu\nu,\mu'\nu'} Y_{\mu'\nu'} \right],
    \nonumber
    \\
    \delta H^{02}_{\mu\nu} & = -(E_\mu+E_\nu)Y_{\mu,\nu}
    + \sum_{\mu'<\nu'} \left[ B^*_{\mu\nu,\mu'\nu'} X_{\mu'\nu'} +  A^*_{\mu\nu,\mu'\nu'} Y_{\mu'\nu'} \right],
\end{align}
and assume that the residual interaction is relatively weak, i.e., 
$|\delta H^{20}_{\mu\nu}| \ll 1$ and $|\delta H^{02}_{\mu\nu}| \ll 1$ in addition to 
ground-state correlations being small ($||Y||_F \ll 1$), the QRPA matrix can be 
simplified to
\begin{align}
        A_{\mu\nu,\mu'\nu'} &\approx (E_\mu+E_\nu)\delta_{\mu\nu,\mu'\nu'}, \\
        B_{\mu\nu,\mu'\nu'} &\approx 0,
\end{align}
which is known as the cranking approximation. In this case, the QRPA eigenproblem 
\eqref{Eq:QRPAdiag1} has a trivial solution: $X=\boldsymbol{I}, Y=\boldsymbol{0}$, 
and all the relative errors $\varepsilon_i$ are identically equal to zero. If 
instead we assume $B_{\mu\nu,\mu'\nu'}\approx 0$ while retaining the full matrix 
$A$, then the QRPA eigenproblem \eqref{Eq:QRPAdiag1} again has a trivial 
solution: $X=Q, Y=\boldsymbol{0}$, where the unitary matrix 
$Q\in\mathbb{C}^{N_p\times N_p}$ diagonalizes the Hermitian matrix: 
$A=Q\Lambda Q^\dagger$. The relative errors $\varepsilon_i$ are then again zero. 
Because $A$ has dominant diagonal elements $E_\mu+E_\nu$, we can expect that the 
eigenvectors $Y^{i}_{\mu,\nu}$ will be generally small in magnitude, except for 
a few modes with small eigenfrequencies $\Omega_i$ for which the diagonal elements 
$E_\mu+E_\nu$ are relatively small. This is the reason why we can \emph{a priori} 
expect very small relative errors $\varepsilon_i$, for all $i=1,\dots,N_p$, 
except for a few low-lying modes, and, therefore, why the introduced estimator 
$\rhohat$ is believed to be a sufficiently good approximation of $\rho(\omega)$.


\section{Practical aspects}
\label{sec:practical}


\subsection{Sampling the level density estimator}

In practice, random variables can only be sampled, and the expected values in 
Eq.~\eqref{Eq:LevelDensityEstimatorDefinition} can be estimated by computing 
average values over a collection of samples. If we generate a collection of 
$N_s$ samples of complex vectors $F_{i_s}^{20},F_{i_s}^{02}\in\mathbb{C}^{N_p}$, 
for $i_s=1,\dots,N_s$, such that their real and imaginary parts are independently 
drawn from a Gaussian distribution with zero \nicolas{expected} value and variance $\sigma^2$ 
as in Eq. \eqref{Eq:F20F02randomVariable}, and if for each of those generated 
excitation operators $\hat{F}_{i_s}$ we compute the response function 
$dB/d\omega(\omega,\hat{F}_{i_s})$ on the positive frequencies domain 
$\omega\geq 0$, then we can approximate the level density estimator 
\eqref{Eq:LevelDensityEstimatorDefinition} as
\begin{equation}\label{Eq:LevelDensityFromSample}
    \rhohat =
    N_p
    \ddfrac
    {\frac{1}{N_s}\sum_{i_s=1}^{N_s} \phantom{\int_{0}^{+\infty}} \frac{dB(\omega\geq 0,\hat{F}_{i_s})}{d\omega} \phantom{d\omega'}}
    {\frac{1}{N_s}\sum_{i_s=1}^{N_s} \int_{0}^{+\infty} \frac{dB(\omega'\geq 0,\hat{F}_{i_s})}{\review{d\omega'}} d\omega' }
    .
\end{equation}
Notice that the denominator in the previous equation is an average over all zeroth 
moments $m_0(\hat{F}_{i_s})$. Of course, the larger the sample size $N_s$, the 
better the approximation but the approximation of the level density function 
does not depend on the variance $\sigma^2$. In principle, one can select any 
value; in practice, a reasonable choice is to take $\sigma \approx 1$, in 
whichever measuring unit we express the excitation operator $\hat{F}$.  
It is worth mentioning that \eqref{Eq:epsiDefinition} and \eqref{Eq:relErrorAtOmegai}
hold in the $N_s\to+\infty$ limit where the expected value matches the computed
average value from the collection of samples.

\subsection{Kernel Polynomial Method}


\review{
The estimator \eqref{Eq:LevelDensityFromSample} seems very practical at first, 
since any method which can evaluate the response function $dB/d\omega(\omega,\hat{F})$
for an arbitrary excitation operator $\hat{F}$,
as well
as the integral in the denominator of Eq. \eqref{Eq:LevelDensityFromSample} is applicable.
For example, one can solve the classical point-by-point FAM equations
to compute the numerator on a fine $\omega$ grid with small smearing $\gamma$,
or use the iterative Arnoldi method from Ref.~\cite{PhysRevC.81.034312}.
The denominator can be evaluated by
employing a contour integration technique, as presented in Ref.~\cite{hinohara2013lowenergy}.
However this calculation
needs to be performed both over a very large $\omega$ domain with small smearing $\gamma$ and for sufficiently large sample size $N_s$. As a result, the 
computational cost of the method could become prohibitive.
This issue can be largely mitigated by employing the Kernel Polynomial Method (KPM) introduced in Ref.~\cite{bjelcic2022chebyshev}. The KPM is specifically designed to efficiently compute the 
response function over the entire $\omega\geq 0$ region of interest 
for arbitrary excitation operator.
In the following we briefly introduce the KPM method and apply it to evaluate the expression in Eq. \eqref{Eq:LevelDensityFromSample}.
}

We define the bounding frequency $\Omega_b$, such that all the QRPA eigenmodes 
are contained within the interval: $\Omega_i\subseteq \langle 0,+\Omega_b \rangle$.
The main idea of the KPM method is to expand the response function in Chebyshev series
\begin{equation}\label{Eq:KPMresponseFunction}
    \frac{dB^{\mathrm{(KPM)}}(\omega,\hat{F})}{d\omega} = 
    \frac{2/\pi}{\Omega_b\sqrt{1-(\omega/\Omega_b)^2}} \sum_{n=0}^{2N_{\mathrm{it}}} \mu_n(\hat{F}) T_n\left(\frac{\omega}{\Omega_b}\right),
\end{equation}
where the coefficients $\{\mu_n(\hat{F})\}_{n=0}^{2N_{\mathrm{it}}}$ are 
computed via the recursive relations given in Ref.~\cite{bjelcic2022chebyshev}. 
The computation of coefficients $\{\mu_n(\hat{F})\}_{n=0}^{2N_{\mathrm{it}}}$
requires exactly $N_{\mathrm{it}}$ evaluations of the QRPA mapping 
\eqref{Eq:QRPAmapping}. In order to damp the Gibbs oscillations, one needs to 
multiply these coefficients with the kernel coefficients 
$\left(g_n^{(2N_{\mathrm{it}}+1)}\right)_{n=0}^{2N_{\mathrm{it}}}$: 
$\mu_n(\hat{F})\longleftarrow g_n^{(2N_{\mathrm{it}}+1)} \mu_n(\hat{F})$. In 
this paper, we use the Jackson kernel \cite{bjelcic2022chebyshev} unless stated 
otherwise. Then, the obtained KPM response function is approximately equal to
\begin{align}
    \frac{dB^{\mathrm{(KPM)}}(\omega,\hat{F})}{d\omega}
    \approx
    &
    +\sum_{i=1}^{N_p} |\langle i | \hat{F} | 0 \rangle |^2 \delta_{\sigma_{\mathrm{KPM}}}(\omega-\Omega_i)
    \nonumber
    \\
    &
    -
    \sum_{i=1}^{N_p} |\langle 0 | \hat{F} | i \rangle |^2 \delta_{\sigma_{\mathrm{KPM}}}(\omega+\Omega_i),
\end{align}
where $\delta_{\sigma_{\mathrm{KPM}}}(\omega)$ is the Gaussian function: 
\begin{equation}
    \delta_{\sigma_{\mathrm{KPM}}}(\omega) = \frac{1}{\sqrt{2\pi}{\sigma_{\mathrm{KPM}}}}\exp\left(-\frac{1}{2}\frac{\omega^2}{\sigma^2_{\mathrm{KPM}}}\right)
\end{equation}
of width ${\sigma_{\mathrm{KPM}}}$
\begin{equation}\label{Eq:sigmaKPM}
    {\sigma_{\mathrm{KPM}}} = \Omega_b \frac{\pi}{2N_{\mathrm{it}}+1}.
\end{equation}
The number $N_{\mathrm{it}}$ of KPM iterations is determined by the desired 
resolution ${\sigma_{\mathrm{KPM}}}$ according to Eq.\eqref{Eq:sigmaKPM}. In 
the limit $N_{\mathrm{it}} \rightarrow +\infty$, we have indeed: 
$dB^{(\mathrm{KPM})}/d\omega \rightarrow dB/d\omega $. It is not difficult to 
see that the zeroth moment $m_0(\hat{F})$ of \eqref{Eq:KPMresponseFunction} is
\begin{equation}
    m_0(\hat{F}) = \int_{0}^{+\Omega_b} \frac{dB^{\mathrm{(KPM)}}(\omega',\hat{F})}{\review{d\omega'}} d\omega' = \sum_{n=0}^{2N_{\mathrm{it}}} \mu_n(\hat{F}) \frac{\sin\left(\frac{n\pi}{2}\right)}{\left(\frac{n\pi}{2}\right)}.
\end{equation}

In order to evaluate the level density estimator $\rhohat$ given in 
Eq.~\eqref{Eq:LevelDensityFromSample}, one runs the KPM calculation for a 
collection of $N_s$ random operators $\{\hat{F}_{i_s}\}_{i_s=1}^{N_s}$. After 
obtaining the collection of coefficients: 
$\big\{ ( \mu_n(\hat{F}_{i_s}) )_{n=0}^{2N_{\rm it}} \big\}_{i_s=1}^{N_s}$, the 
QRPA level density estimator is given by
\begin{equation}\label{Eq:FinalKPMLevelDensityEstimator}
    \rhohat = N_p
    \phantom{.}
    \ddfrac
    { \sum_{n=0}^{2N_{\mathrm{it}}}
    \overline{\mu_n(\hat{F}_{i_s})}
    \phantom{.}
    \frac{2/\pi}{\Omega_b\sqrt{1-(\omega/\Omega_b)^2}}T_n\left(\frac{\omega}{\Omega_b}\right)}
    {\sum_{n=0}^{2N_{\mathrm{it}}}
    \overline{\mu_n(\hat{F}_{i_s})}
    \phantom{.}
    \frac{\sin\left(\frac{n\pi}{2}\right)}{\left(\frac{n\pi}{2}\right)} 
    \phantom{............................}
    },
\end{equation}
where $\overline{\mu_n(\hat{F}_{i_s})}$ are the averaged coefficients over 
the collection of randomly generated samples
\begin{equation}\label{Eq:AveragedKPMcoefficients}
    \overline{\mu_n(\hat{F}_{i_s})}
    =
    \frac{1}{N_s}\sum_{i_s=1}^{N_s} \mu_n(\hat{F}_{i_s}),  \quad \forall n=0,1,\dots,2N_{\mathrm{it}}.
\end{equation}
The KPM computation for each $i_s$ is independent from the others and therefore 
can be run fully in parallel.


\subsection{Spurious modes removal}

In practical calculations, spurious modes, also know as the Nambu-Goldstone modes \cite{hinohara2015collective}, 
often contaminate the response function due to their high transition probabilities 
at positive near-zero frequencies. In most nuclei, the spurious pairing $K^\pi=0^+$, 
rotational $K^\pi=1^+$ and translational $K^\pi=0^-,1^-$ modes in the response 
function will result in a corresponding spurious contribution to the calculated 
level density function. 

The idea is to eliminate the contribution of spurious modes in the low-energy region by
shifting them to high-energy region.
Once the spurious mode with eigenfrequency $\Omega_1>0$ and eigenvector 
$x_1,y_1\in\mathbb{C}^{N_p}$ is identified, the only thing that needs to be 
modified is the QRPA mapping \eqref{Eq:QRPAmapping} which now becomes
\begin{multline}\label{Eq:QRPASpuriousMappingModified}
    \begin{bmatrix} x \\ y \end{bmatrix}
    \mapsto
    \begin{bmatrix} A & B \\ B^* & A^*  \end{bmatrix}
    \begin{bmatrix} x \\ y \end{bmatrix} \\
    +
    (\widetilde{\Omega}_1-\Omega_1)
    \left\{ 
    \left( x_1^\dagger x - y_1^\dagger y \right)
    \begin{bmatrix} +x_1 \\ -y_1 \end{bmatrix} +
    \left( x_1^T y - y_1^T x \right)
    \begin{bmatrix}-y_1^* \\ +x_1^* \end{bmatrix}
    \right\},
\end{multline}
where $\widetilde{\Omega}_1>0$ is the position of the shifted spurious mode. In 
practice most spurious modes have $ \Omega_1 < 0.5 \text{ }\mathrm{MeV}$. By 
choosing large values of $\widetilde{\Omega}_1>0$, e.g. 
$\widetilde{\Omega}_1 \geq 200$ MeV, one can shift these modes to arbitrarily 
high energy. Additional details on the practical implementation of this technique 
are given in  \ref{Appendix:Spurious}.


\begin{algorithm}[!htb]
\footnotesize
    \SetAlgoLined
    \caption{QRPA level density with FAM+KPM}
\phantom{.}

    \textbf{Input:}
    \parbox{0.44\textwidth}{
    \begin{itemize}
        \item Mapping:
        $\left[ \begin{smallmatrix} x \\ y \end{smallmatrix} \right]\in\mathbb{C}^{2N_p}
        \mapsto
        \left[ \begin{smallmatrix} A & B \\ B^* & A^* \end{smallmatrix} \right]
        \left[ \begin{smallmatrix} x \\ y \end{smallmatrix} \right]\in\mathbb{C}^{2N_p},$ 
        provided by a FAM solver.
        \item Width $\sigma_{\mathrm{KPM}}$ of Gaussian approximation of Dirac delta 
        function.
        \item Bounding frequency $\Omega_b$: $\forall\Omega_i, \Omega_i 
        \in \langle 0 , +\Omega_b \rangle $.
        \item Optionally provide: spurious mode eigenfrequency $\Omega_1>0$, spurious 
        mode eigenvector $x_1,y_1\in\mathbb{C}^{N_p}$, and a desired position of the 
        shifted spurious eigenfrequency $\widetilde{\Omega}_1$ such that 
        $\widetilde{\Omega}_1 \in  \langle 0 , +\Omega_b \rangle$.
        How to obtain $\{\Omega_1,x_1,y_1\}$ is  explained in \ref{Appendix:Spurious}.
        \item Number of samples $N_s$ in a collection of randomly generated \\
        excitation operators  $\{\hat{F}_{i_s}\}_{i_s=1}^{N_s}$.
    \end{itemize}
    }

\For{ $i_s=1,2,\dots,N_{s}$ }
{
\parbox{0.425\textwidth}{
\begin{itemize}
   \item Generate a random operator $\hat{F}_{i_s}$ defined by vectors: 
   $\operatorname{Re}[F^{20}_{i_s}], \operatorname{Re}[F^{02}_{i_s}], 
   \operatorname{Im}[F^{20}_{i_s}], \operatorname{Im}[F^{02}_{i_s}]\in\mathbb{R}^{N_p}$, 
   where each  component is drawn independently from $\mathcal{N}(0,\sigma^2)$.
   \item Compute coefficients $\mu_n(\hat{F}_{i_s})$ using spurious-corrected mapping \eqref{Eq:QRPASpuriousMappingModified} as described in Ref.~\cite{bjelcic2022chebyshev}.
   The number $N_{\mathrm{it}}$ of  FAM iterations needed is computed from Eq.~\eqref{Eq:sigmaKPM}.
    \item  Apply the kernel transformation: $\mu_n(\hat{F}_{i_s})\longleftarrow g_n^{(2N_{\mathrm{it}}+1)} \mu_n(\hat{F}_{i_s})$, \\
    with the Jackson kernel coefficients 
    $g_n^{(2N_{\mathrm{it}}+1)}$ given in Ref. \cite{bjelcic2022chebyshev}.
\end{itemize}
}
}

\parbox{0.45\textwidth}{
\begin{itemize}
    \item Compute the averaged coefficients $\overline{\mu_n(\hat{F}_{i_s})}$ as 
    in Eq.~\eqref{Eq:AveragedKPMcoefficients}.
    \item Using Eq.~\eqref{Eq:FinalKPMLevelDensityEstimator}, evaluate $\rhohat$ 
    on a grid of frequencies $\omega\geq 0$, e.g., with the FFT-based method of Ref.~\cite{bjelcic2022chebyshev}.
\end{itemize} 

\textbf{Output:}
\begin{itemize}
    \item QRPA level density estimator $\rhohat$ evaluated on a $\omega\geq 0$ grid.
\end{itemize}
}
\label{Algorithm:Random}
\end{algorithm}

\subsection{Algorithm}

Algorithm \ref{Algorithm:Random} summarizes the main steps of our method. The total 
number $\#_{\mathrm{QFAM}}$ of FAM iterations required is
\begin{equation}\label{Eq:Nqfam}
    \#_{\mathrm{QFAM}} = N_s N_{\mathrm{it}} = N_s \ceil[\bigg]{\frac{1}{2}\left( \frac{\Omega_b\pi}{\sigma_{\mathrm{KPM}}} - 1 \right)},
\end{equation}
where $N_{\mathrm{it}}$ is obtained from Eq. \eqref{Eq:sigmaKPM}.
We emphasize again that the loop in Algorithm \ref{Algorithm:Random} is 
naturally parallelizable. For example, in the case of a non-relativistic FAM 
solver, if one picks $\Omega_b\approx 250 $ MeV and 
$\sigma_{\mathrm{KPM}} = 0.05$ MeV, it takes around $N_{\mathrm{it}}\approx 8000$ 
FAM iterations per sample. If we assume that each FAM iteration lasts around 
$\approx 5$ s, and that we \nicolas{can launch} $N_s \approx 200$ \nicolas{tasks in parallel}, 
one can expect to compute the level density function in $\approx 11$ hours, 
which makes the method feasible in practice.


\section{Numerical tests}
\label{sec:tests}

In this section, we show a series of numerical tests to evaluate the performance 
of the new method. We first compute the QRPA level density for synthetically 
generated QRPA-like matrices to illustrate the convergence of the method as we
increase the number of samples $N_s$. We then use the open-source code \texttt{skyrme\_rpa} 
\cite{colo2013selfconsistent} to solve the RPA equation in heavy spherical nuclei 
and benchmark the QRPA level density of specific $K^{\pi}$ modes.


\subsection{Examples using synthetically generated QRPA matrix}

In this section we synthetically generate QRPA-like matrices $A$ and $B$ by 
employing the procedure described in Appendix B of Ref. \cite{bjelcic2022chebyshev}.
The matrix dimension and the bounding frequency are set to $N_p=200$ and 
$\Omega_b=250$ MeV, respectively. 100 random eigenfrequencies $\Omega_i$ are 
generated uniformly in the range from 0 MeV to 200 MeV and combined with another 
100 random eigenfrequencies $\Omega_i$ generated uniformly in the range from
0 MeV to 30 MeV, with the intent of making the resulting QRPA spectrum more 
dense in the low energy region.

Next, we generate a uniformly random sequence of values $\theta_1,\dots,\theta_{N_p}$ 
in the interval $[ 0 , \theta_{\mathrm{max}} ]$, and two unitary matrices 
$C,D\in\mathbb{C}^{N_p\times N_p}$ as Q factors in the QR decomposition of two 
random $N_p\times N_p$ complex matrices. The $X$ and $Y$ matrices are constructed 
as
\begin{equation}
    X = D \operatorname{diag}\left[\cosh\theta_i\right]_{i=1}^{N_p} C
    \quad
    \text{and}
    \quad
    Y = D^* \operatorname{diag}\left[\sinh\theta_i\right]_{i=1}^{N_p} C,
\end{equation}
and then used to generate the QRPA matrices $A$ and $B$ as
\begin{equation}
    A = +X\Omega X^\dagger +
    (Y\Omega Y^\dagger)^*
    \quad
    \text{and}
    \quad
    B = -X\Omega Y^\dagger - (X\Omega Y^\dagger)^T.
\end{equation}
Note that the parameter $\theta_{\mathrm{max}}$ determines the magnitudes of 
$||Y\boldsymbol{e}_i||^2$ and therefore the corresponding theoretical relative 
error $\varepsilon_i$ in the limit $N_s\to +\infty$, according to 
Eq.\eqref{Eq:epsiDefinition}. 

\begin{figure}[!htb]
\centering
\includegraphics[width=0.95\linewidth]{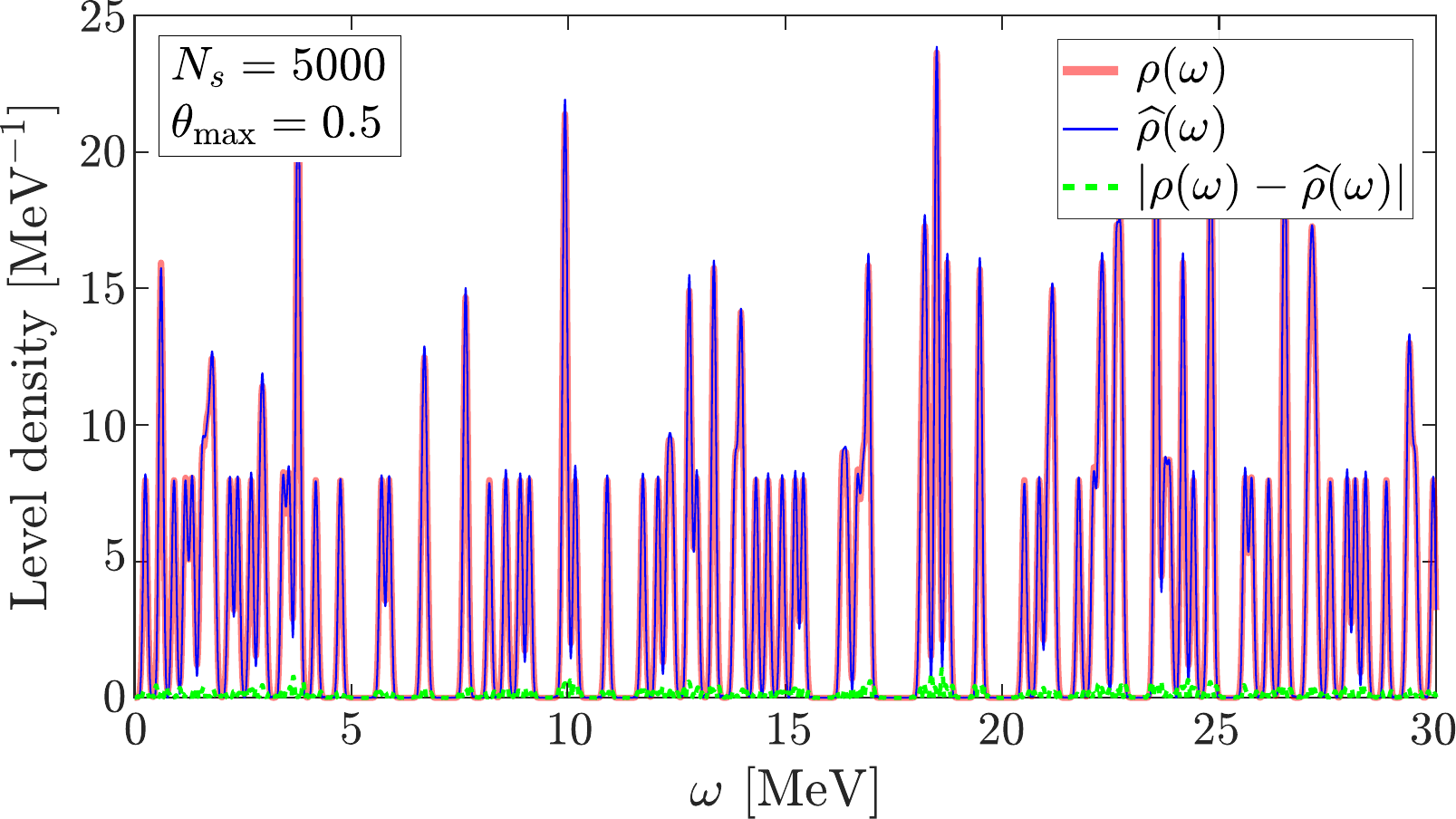}
\caption{
Comparison of the level 
density function $\rho(\omega)$ and its estimator $\rhohat$ computed for a very large number of samples
$N_s=5000$  and $\theta_{\mathrm{max}}=0.5$.
$\rho(\omega)$ is calculated by approximating the Dirac delta function
with a Gaussian function, using a standard deviation of $\sigma_{\mathrm{KPM}}$.
}
\label{Fig:qrpa_ld_synthetic}
\end{figure}

We set the width for the Gaussian approximation of 
the Dirac delta function equal to $\sigma_{\mathrm{KPM}}=0.05$ MeV, which for 
$\Omega_b=250$ MeV according to Eq.~\eqref{Eq:Nqfam} corresponds to the number 
$N_{\mathrm{it}} \approx 8000$ of QFAM iterations needed in the KPM method. We 
draw the components of vectors $F_{i_s}^{20},F_{i_s}^{02}\in\mathbb{C}^{N_p}$ 
according to Eq.~\eqref{Eq:F20F02randomVariable} with a standard deviation of 
$\sigma=1$ MeV. Recall that the error on the estimator $\rhohat$ is independent 
of the value of $\sigma$.
Figure \ref{Fig:qrpa_ld_synthetic} shows the level 
density $\rho(\omega)$ and its estimator $\rhohat$ computed for a very large number of samples
$N_s=5000$  and $\theta_{\mathrm{max}}=0.5$. 

\subsubsection{Example with fixed $\theta_{\mathrm{max}}$ and varying $N_s$}

\begin{figure}[!htb]
\centering
\includegraphics[width=0.98\linewidth]{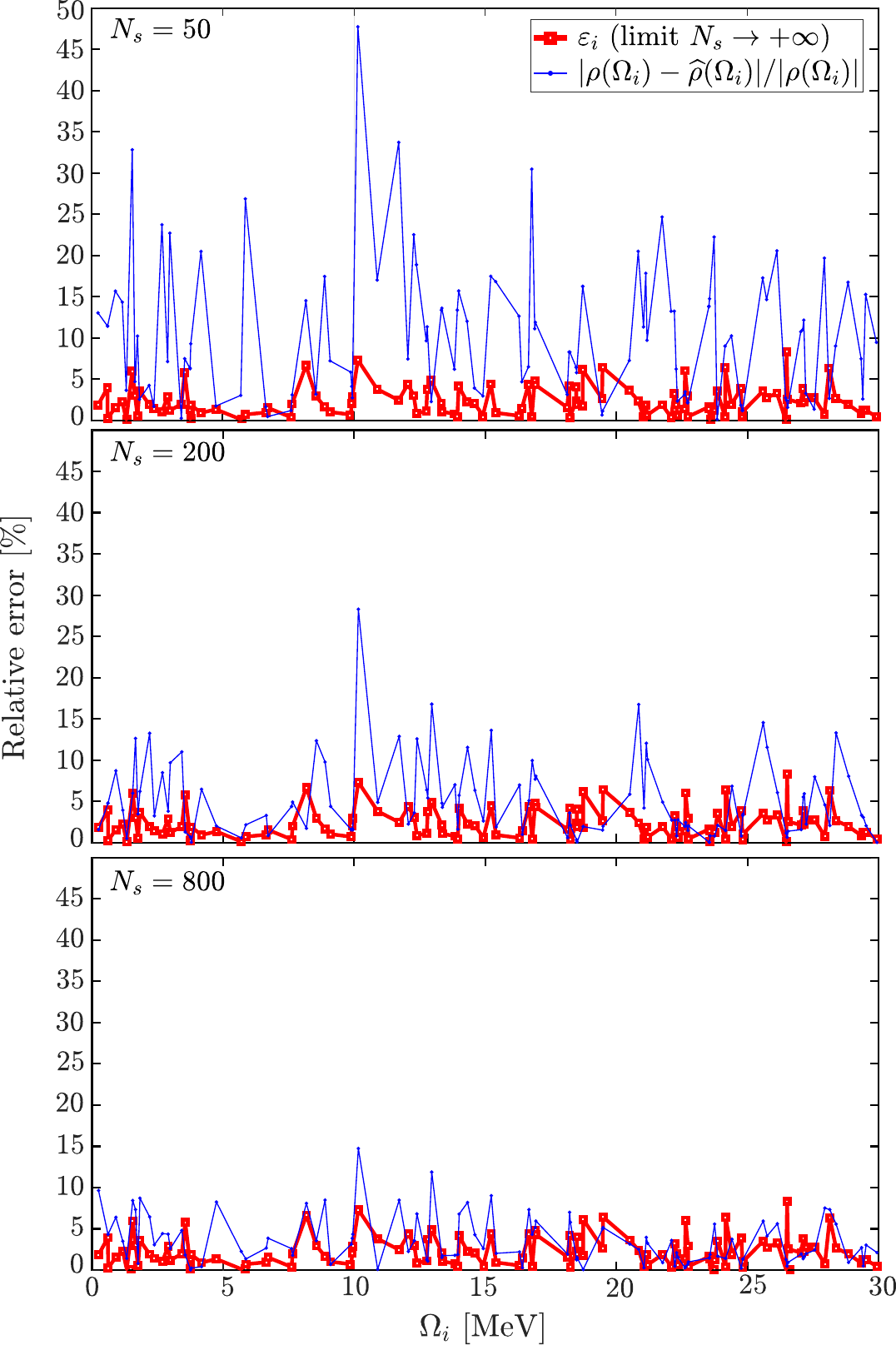}
\caption{The blue curve 
with plain circles shows a relative error
evaluated at eigenfrequencies $\omega=\Omega_i$
between the level density $\rho(\omega)$ and its estimator
$\rhohat$ computed using 
$N_s=50,200$ and $800$ randomly generated excitation operators and 
$\theta_{\rm max} = 1$.
The red curve with open squares shows the theoretical 
relative error $\varepsilon_i$ in the $N_s\to +\infty$ limit.}
\label{Fig:VaryNs}
\end{figure}

In this subsection we fix the value $\theta_{\mathrm{max}} = 1$ and vary the 
number of randomly generated excitation operators $N_s=50$, 200, 800. In 
Fig.~\ref{Fig:VaryNs} we show the
obtained relative error
evaluated at QRPA eigenfrequencies $\omega=\Omega_i$
between the level density estimator $\rhohat$ 
compared to the true level density function $\rho(\omega)$
when 
the Dirac delta function in Eq. \eqref{Eq:LevelDensityDefinition} is replaced by a Gaussian with standard 
deviation of $\sigma_{\mathrm{KPM}}$.
Figure \ref{Fig:VaryNs} also shows the 
theoretical relative error limit $\varepsilon_i$ in the case of an infinitely 
large collection of random excitation operators $N_s\rightarrow +\infty$ as 
defined in Eq. \eqref{Eq:epsilon_i}.
We can see that for $N_s = 800$, the obtained 
relative error is of the same order $\approx 5-10 \%$ as the theoretical limit. 
Therefore, if $\theta_{\mathrm{max}}\approx 1$, a few hundreds of randomly 
generated excitation operators are enough to give an estimate up to 
$\approx 10\%$ relative error.
We remind the reader that \eqref{Eq:epsiDefinition} and \eqref{Eq:relErrorAtOmegai} hold in the limit of
an infinitesimally small width of Dirac delta functions and
the $N_s\to +\infty$ limit. When using a finite-width approximation of the Dirac delta function and a finite number of samples $N_s$, these relations are only approximately valid.

\subsubsection{Example with fixed $N_s$ and varying $\theta_{\mathrm{max}}$}

\begin{figure}[!htb]
\centering
\includegraphics[width=0.98\linewidth]{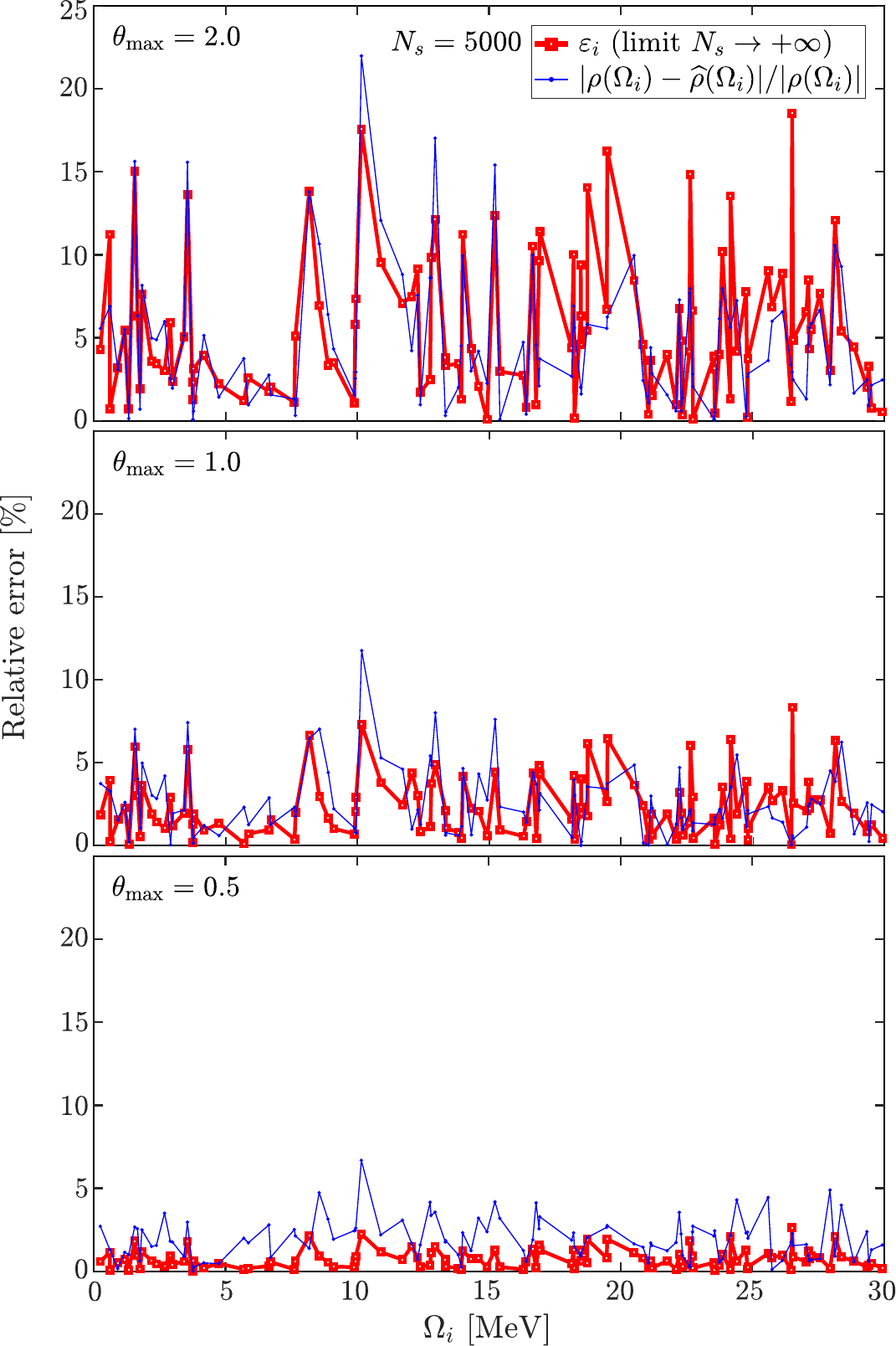}
\caption{
The blue curve 
with plain circles shows a relative error
evaluated at eigenfrequencies $\omega=\Omega_i$
between the level density $\rho(\omega)$ and its estimator
$\rhohat$ computed using a very large number $N_s=5000$
of randomly generated excitation operators and 
$\theta_{\rm max} = 0.5, 1.0, 2.0$.
The red curve with open squares shows the theoretical 
relative error $\varepsilon_i$ in the $N_s\to +\infty$ limit.
}
\label{Fig:VaryThetamax}
\end{figure}

In this subsection we fix the number of samples $N_s$ to a very large value 
$N_s=5000$ and vary the parameter $\theta_{\mathrm{max}}=0.5$, 1.0, 2.0. 
\nicolas{Recall that small values of $\theta_{\rm max}$ correspond to small values 
of the matrix elements of $Y$, hence a smaller error $\varepsilon_i$ owing to Eq.~\eqref{Eq:epsilon_i}
}
Like 
in the previous subsection, in Fig.~\ref{Fig:VaryThetamax} we compare the level 
density estimator $\rhohat$ with the true level density $\rho(\omega)$
by showing
the theoretical relative error 
limit $\varepsilon_i$ and the computed relative error $|\rhohat-\rho(\omega)|/|\rho(\omega)|$ at $\omega=\Omega_i$. 
We first notice that as the value of $\theta_{\mathrm{max}}$ is reduced, the 
theoretical relative error limit $\varepsilon_i$ is reduced \nicolas{indeed}. Secondly, even for 
a large number of samples $N_s=5000$, the computed relative error is primarily 
dominated by statistical errors due to the variations within the finite size of 
the collection of random samples, rather than the theoretical error limit. This 
implies that even under the most favorable condition of $\theta_{\mathrm{max}} = 0$, 
the resulting error would be approximately $5\%$, even with a very large sample 
size of $N_s=5000$.

\subsection{Examples using matrix RPA code \texttt{skyrme\_rpa}}

In this section, we validate the proposed method with the realistic RPA solver 
\texttt{skyrme\_rpa} \cite{colo2013selfconsistent}. This solver is implemented for
closed shell spherical nuclei with Skyrme-type interactions. The Hartree-Fock 
equations are solved on a radial mesh using box boundary conditions and the RPA 
matrix is explicitly constructed and diagonalized for a given value of total angular 
momentum and parity $J^\pi$.

We have modified the \texttt{skyrme\_rpa} code to extract explicitly the 
matrices $A$ and $B$ for given $J^\pi$ with the intent of using them rather
than synthetically generated ones as was done in the previous section.
This shall provide an insight of realistic values for $\Omega_i$ and 
$\theta_i$ which were previously randomly generated. We use the desired width 
$\sigma_{\mathrm{KPM}} =0.05$ MeV, bounding frequency $\Omega_b = 250$ MeV, and 
standard deviation of randomly generated excitation operators $\sigma= 1$ MeV.
Guided by the previous synthetic examples, we choose the number of samples
$N_s = 500$ as a compromise between computational complexity and statistical 
fluctuations.


\subsubsection{$J^\pi = 5^-$ level density of ${}^{120}$Sn}

\begin{figure}[H]
\centering
\includegraphics[width=0.99\linewidth]{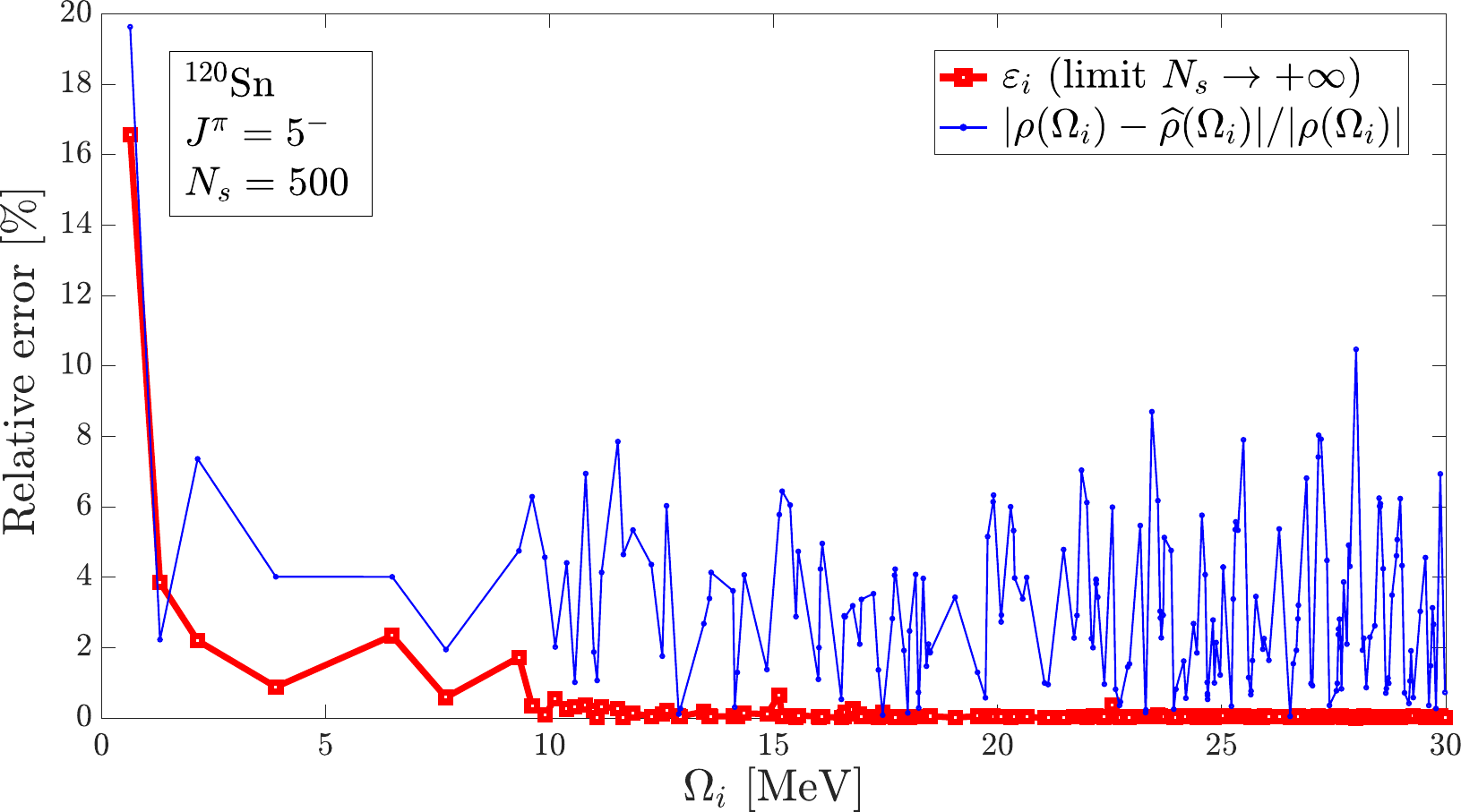}
\caption{Theoretical relative error $\varepsilon_i$ in the limit $N_s\to +\infty$ 
and computed relative error: $|{\rho}(\Omega_i)-\rhohati|/|{\rho}(\Omega_i)|$ for 
$J^\pi=5^-$ states in $^{120}$Sn. Calculations were performed with $N_s=500$ 
randomly generated excitation operators; see text for details about the RPA 
calculation.}
\label{Fig:Sn120}
\end{figure}

We use the same numerical setup as in Section 4.2 of 
Ref.~\cite{bjelcic2022chebyshev}. Calculations were performed by employing the SLy5 
Skyrme interaction on ${}^{120}$Sn in a 20 fm radius box with 0.1 fm radial step 
and 100 MeV cutoff energy; see Ref. \cite{colo2013selfconsistent} for details. The 
resulting RPA matrix for $J^\pi = 5^-$ is of order $N_p = 1310$. First we solve 
the QRPA eigenvalue problem as in Eqs. \eqref{Eq:QRPAdiag1} and \eqref{Eq:QRPAdiag2} 
and obtain the eigenfrequencies $\Omega_i$, and matrices $X,Y$. The values 
of $\theta_1,\dots,\theta_{N_p}$ are obtained by computing the singular values of 
either $X$ or $Y$ matrix.
The maximum angle obtained is equal to: $\operatorname{max}_{i=1,\dots,N_p}\theta_i = 0.38$. As 
demonstrated in the previous section, this a favourable scenario regarding the 
theoretical error $\varepsilon_i$ in the $N_s\to +\infty$ limit. In 
Fig.~\ref{Fig:Sn120} we compare the computed level density estimator against the 
true function. We notice that the obtained relative error is the largest for the 
eigenmode with the lowest eigenfrequency. This is expected because this mode has 
the highest ground-state correlation energy and therefore the largest 
$||Y\boldsymbol{e}_i||^2$, which yields largest $\varepsilon_i$. Other than the 
lowest mode, the obtained relative error is of order $5-10\%$ for other modes, and 
is mostly influenced by statistical fluctuations.

\subsubsection{$J^\pi = 1^-$ level density of ${}^{208}$Pb}

\begin{figure}[!htb]
\centering
\includegraphics[width=0.99\linewidth]{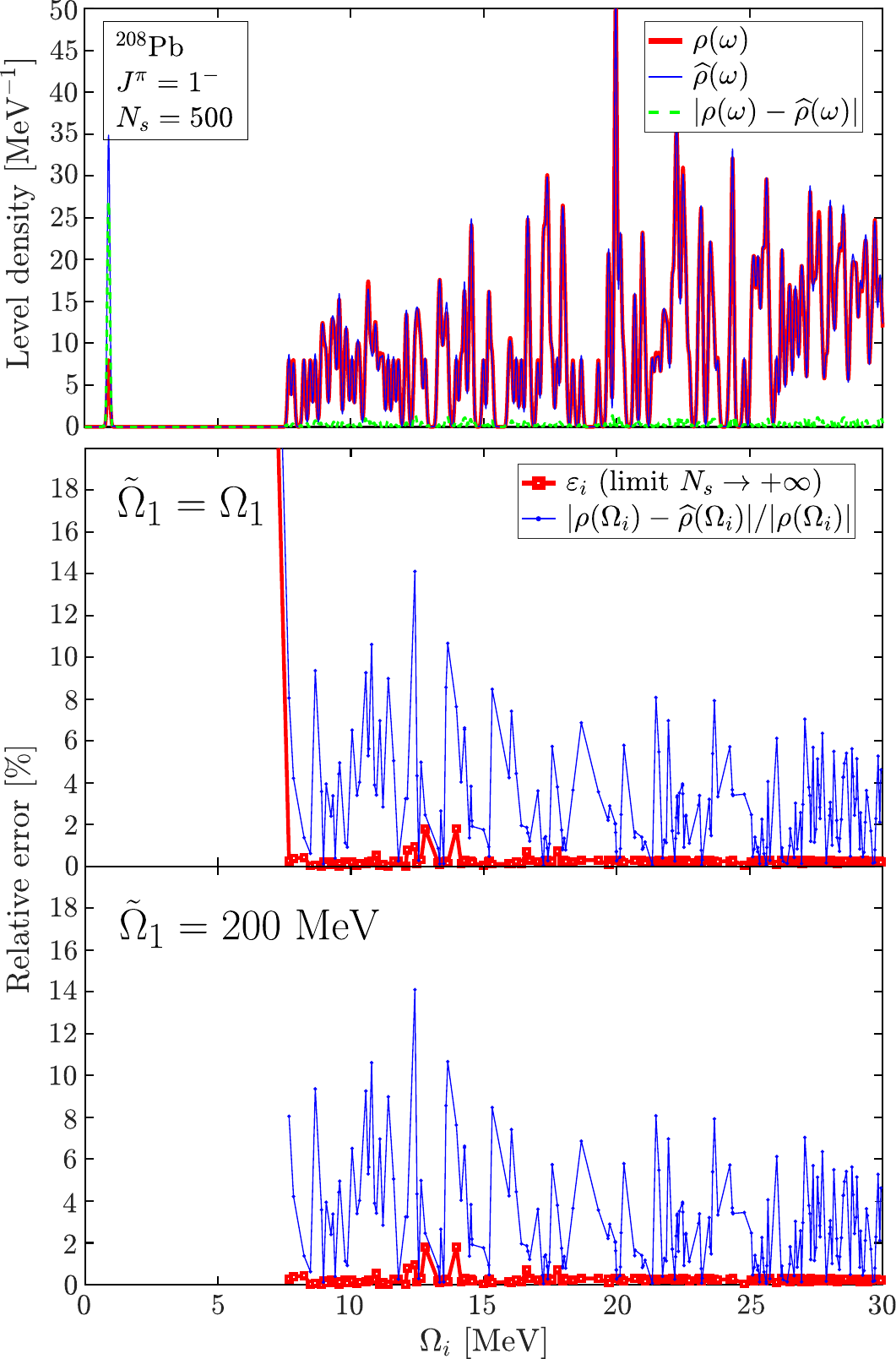}
\caption{Top panel: Level density estimator $\rhohat$, true density $\rho(\omega)$ 
and absolute difference of the two for $J^\pi = 1^-$ states in $^{208}$Pb as a 
function of the energy $\omega$. Calculations were 
performed for $N_{s} = 500$ samples. Middle panel: Relative difference 
between $\rhohat$ and $\rho(\omega)$
and a theoretical relative error limit $\varepsilon_i$ as a 
function of the energy $\Omega_i$ of the RPA eigenstate. Bottom panel: Same as 
middle panel after moving the spurious mode to $\widetilde{\Omega}_{1} = 200$ MeV 
with the procedure described in \ref{Appendix:Spurious}.}
\label{Fig:Pb208}
\end{figure}

In this example we use the SLy5 Skyrme interaction on ${}^{208}$Pb in a box of 
radius 24 fm with 0.1 fm radial step and 100 MeV cutoff energy. The resulting RPA 
matrix for $J^\pi = 1^-$ is of order $N_p=1394$. In this case, the maximum angle 
$\theta_i$ is equal to: $\operatorname{max}_{i=1,\dots,N_p}\theta_i = 1.08$, which 
is why we expect a somewhat larger error than in the previous example. The top panel 
of Fig.~\ref{Fig:Pb208} shows the comparison between the computed level density 
estimator and the correct level density function. The low-lying spurious $1^-$ mode 
is at $\Omega_1 \approx 1$ MeV.

In the middle panel of Fig.~\ref{Fig:Pb208}, we plot the theoretical relative 
error $\varepsilon_i$ in the limit $N_s\to +\infty$ and the computed relative error 
$|{\rho}(\Omega_i)-\rhohati|/|{\rho}(\Omega_i)|$. We notice that the low-lying 
spurious $1^-$ mode yields a very large relative error. This is expected because 
such spurious modes have a very large norm $||Y\boldsymbol{e}_i||^2$ and 
consequently large relative error $\varepsilon_i$ in the $N_s\to +\infty$ limit; see 
Eq.~\eqref{Eq:epsilon_i}. In this particular case, $\varepsilon_i=334\%$ for the 
spurious mode. 

In order to eliminate the contribution of the spurious mode to the level density in 
the low-energy region, we follow the method outlined in 
\ref{Appendix:Spurious} and shift the spurious mode to high energy 
$\widetilde{\Omega}_{1} = 200$ MeV. The bottom panel of Fig.~\ref{Fig:Pb208} shows 
the relative error after this operation. Now the obtained relative error is of order 
$5-10\%$ and is mostly influenced by statistical fluctuations.


\section{Conclusion}

In this work, we have proposed a new method to approximate the QRPA level density, 
that is, the density of vibrational states that result from the diagonalization 
of the QRPA matrix.
Our approach relies on using the Finite Amplitude Method to 
compute the linear responses for a collection of random excitation operators that probe the 
QRPA spectrum.
While the error on the level density decreases with the number of 
random operators, i.e. the number of samples,
we derive a theoretical relative error bound in the case of infinite number of samples.
%
This bound is expected to be relatively small; at the cranking 
approximation, it is in fact exactly zero.
We use the Kernel 
Polynomial Method to efficiently compute the estimator of the QRPA level density. 
We illustrate the convergence of the method with respect to the number of samples 
for various magnitudes of $||Y||_F$ by synthetically generating QRPA-like matrices.
Using publicly available RPA code, we also show that we can approximate 
the RPA level density to within 5--10\%. 

While our method is no substitute to exact diagonalization of the QRPA matrix, it 
offers two main advantages: (i) it is easily applicable to heavy deformed nuclei, 
where the cost of building the QRPA matrix can quickly become prohibitive (ii) 
computationally, it is a naturally parallel problem since the linear response of 
each sample can be computed independently. The next step will be to use such QRPA 
level densities to compute the full level density of the nucleus by folding it 
with rotational bands.

\section{Acknowledgements}
Discussions with Tong Li are warmly acknowledged.
This work was performed under the auspices of the U.S. Department of Energy by Lawrence Livermore National Laboratory under contract DE-AC52-07NA27344.
Lawrence Livermore National Security, LLC.
Computing support came from the Lawrence Livermore National Laboratory (LLNL)
Institutional Computing Grand Challenge program.


\appendix

\section{\label{Appendix:Spurious}Removal of spurious modes}

The main idea stems from the following facts. If $H\in\mathbb{C}^{n\times n}$ is 
a Hermitian matrix with eigenvalues: $\lambda_1,\dots,\lambda_n\in\mathbb{R}$, and 
corresponding eigenvectors: $u_1,\dots,u_n\in\mathbb{C}^n$, 
\begin{equation}
    H = \sum_{i=1}^n \lambda_i u_i u_i^\dagger,
\end{equation}
then for any $\tilde{\lambda}_1\in\mathbb{R}$, the updated rank-1 Hermitian matrix
\begin{equation}
   \widetilde{H} =  H + (\tilde{\lambda}_1-\lambda_1)u_1u_1^\dagger,
\end{equation}
has eigenvalues: $\tilde{\lambda}_1,\lambda_2,\dots,\lambda_n\in\mathbb{R}$ and 
the same corresponding eigenvectors as the matrix $H$: $u_1,u_2,\dots,u_n 
\in\mathbb{C}^n$.This property can be applied to the QRPA matrix $\mathcal{S}$ of 
Eq.~\eqref{Eq:QRPAmatrix}. Let us introduce a shorthand notation for the columns 
of the $X$ and $Y$ matrices of eigenvectors: $x_i=X\boldsymbol{e}_i$ and 
$y_i = Y\boldsymbol{e}_i$, for $i=1,\dots,N_p$. For any positive 
$(\widetilde{\Omega}_i)_{i=1}^{N_p}$, let us define the modified matrices,
\begin{align}
    \widetilde{A} &= A + \sum_{i=1}^{N_p} (\widetilde{\Omega}_i-\Omega_i) \left[ + x_ix_i^\dagger + (y_iy_i^\dagger)^* \right],
\\
    \widetilde{B} &= B + \sum_{i=1}^{N_p} (\widetilde{\Omega}_i-\Omega_i) \left[ - x_iy_i^\dagger - (x_iy_i^\dagger)^T \right].
\end{align}
Then the new QRPA matrix where $A,B$ are replaced by $\widetilde{A},
\widetilde{B}$ still satisfies both Eqs.~\eqref{Eq:QRPAdiag1} and 
\eqref{Eq:QRPAdiag2} with the same eigenvectors $X$ and $Y$ only with 
eigenfrequencies $\widetilde{\Omega}_i$ instead of $\Omega_i$. The QRPA 
mapping \eqref{Eq:QRPAmapping} acquires the new form
\begin{multline}\label{eq:new_mapping}
    \begin{bmatrix} x \\ y \end{bmatrix}
    \mapsto
    \begin{bmatrix} \widetilde{A} & \widetilde{B} \\ \widetilde{B}^* & \widetilde{A}^* \end{bmatrix}
    \begin{bmatrix} x \\ y \end{bmatrix}     =
    \begin{bmatrix} {A} & {B} \\ {B}^* & {A}^* \end{bmatrix}
    \begin{bmatrix} x \\ y \end{bmatrix}\\
    +
    \sum_{i=1}^{N_p}
    (\widetilde{\Omega}_i-\Omega_i)
    \left\{
    \left( x_i^\dagger x - y_i^\dagger y \right)
    \begin{bmatrix} +x_i \\ -y_i \end{bmatrix} +
    \left( x_i^T y - y_i^T x \right)
    \begin{bmatrix} -y_i^* \\ +x_i^* \end{bmatrix}
    \right\} .
\end{multline}
In practice, we often want to shift only a single spurious mode $\Omega_1>0$ 
from near-zero value to some other value $\widetilde{\Omega}_1>0$, away from 
the low-energy region we are typically interested in. If we somehow manage to 
obtain the eigenfrequency $\Omega_1>0$ of this spurious mode $\Omega_1>0$, 
then we can simply apply the formula \eqref{eq:new_mapping} with $N_p = 1$ to 
obtain exactly the same response function but with shifted spurious mode 
$\Omega_1\rightarrow \widetilde{\Omega}_1$.

The question is therefore how to obtain the low-lying QRPA modes $\Omega_i$ and 
the corresponding $(x_i,y_i)$, $i=1,\dots,k$, for a subset of $k$ modes. There 
are efficient methods for computing the $k$ eigenvalues with the smallest norm 
of a general matrix $H\in\mathbb{C}^{n\times n}$ which only require the inverse
mapping: $x\mapsto H^{-1}x$. One such method is included in ARPACK \cite{arpack}, a 
numerical software library for solving large scale eigenvalue problems using the 
implicitly restarted Arnoldi method.

Consider the mapping
\begin{equation}\label{Eq:QRPAinverseMapping}
    \begin{bmatrix} f^{20} \\ f^{02} \end{bmatrix}
    \mapsto
    \left(
    \begin{bmatrix} +\boldsymbol{I} & \boldsymbol{0} \\ \boldsymbol{0} & -\boldsymbol{I} \end{bmatrix}
    \begin{bmatrix} {A} & {B} \\ {B}^* & {A}^* \end{bmatrix}
    \right)^{-1}
    \begin{bmatrix} f^{20} \\ f^{02} \end{bmatrix}.
\end{equation}
According to Eq.~\eqref{Eq:LinearResponseEquation}, the mapping 
\eqref{Eq:QRPAinverseMapping} corresponds to solving the linear response 
equation for zero complex frequency $\omega_\gamma=0$ and excitation operator 
given by $F^{20}=-f^{20}$ and $F^{02}=+f^{02}$. This linear response equation is 
precisely what the FAM solves. Therefore, with the help of e.g. ARPACK, one can 
compute the $2k$ eigenvalues with the smallest norm: 
$\pm\Omega_1,\dots,\pm\Omega_k$, and the corresponding eigenvectors for the 
following matrix
\begin{equation}
    \left[\begin{matrix}
        +\boldsymbol{I} & \boldsymbol{0} \\ \boldsymbol{0} & -\boldsymbol{I}
    \end{matrix}\right]
    \left[\begin{matrix}
        {A} & {B} \\ {B}^* & {A}^*
    \end{matrix}\right]
    \left[\begin{matrix}
        u_i \\ v_i
    \end{matrix}\right]
    =
    \Omega_i
        \left[\begin{matrix}
        u_i \\ v_i
    \end{matrix}\right], \text{ for } i=1,\dots,k.
\end{equation}
Recall that the eigenvalues of the QRPA matrix $\mathcal{S}$ of 
Eq.~\eqref{Eq:QRPAmatrix} come in pairs $(+\Omega_i,-\Omega_i)$ with respective 
eigenvectors
\begin{equation}
   +\Omega_i: \begin{bmatrix} u_i \\ v_i \end{bmatrix}, \qquad
   +\Omega_i: \begin{bmatrix} v_i^* \\ u_i^* \end{bmatrix}.
\end{equation}
Eigensolvers such as ARPACK typically normalize the eigenvector such that 
$||u_i||^2+||v_i||^2 = 1$. We define the eigenvectors
\begin{equation}\label{Eq:QRPAeigenvectorsWithARPACK}
    \begin{bmatrix} x_i \\ y_i \end{bmatrix} =
    \frac{1}{\sqrt{||u_i||^2 - ||v_i||^2}}
    \begin{bmatrix} u_i \\ v_i \end{bmatrix}, \text{ for } i=1,\dots,k,
\end{equation}
which are normalized with respect to the QRPA metric: $||x_i||^2 - ||y_i||^2 = 1$. 
Mathematically spurious zero-energy modes are actually non-normalizable 
\cite{hinohara2015collective}, because for them $||u_i||^2 - ||v_i||^2 = 0$. 
In practice, however, spurious modes never have exactly vanishing eigenvalue and one 
can normalize them. 

In the case where the QRPA eigenmodes are non degenerate, two eigenvectors 
\eqref{Eq:QRPAeigenvectorsWithARPACK} for $i\neq j$, hence $\Omega_i \neq \Omega_j$,
are orthogonal with respect to the QRPA metric since it is easy to show that
\begin{equation}
    \Omega_j
    \begin{bmatrix} x_i \\ y_i \end{bmatrix}^\dagger
    \begin{bmatrix} + \boldsymbol{I} & \boldsymbol{0} \\ \boldsymbol{0} & -\boldsymbol{I} \end{bmatrix}
    \begin{bmatrix} x_j \\ y_j \end{bmatrix} =
    \Omega_i
    \begin{bmatrix} x_i \\ y_i \end{bmatrix}^\dagger
    \begin{bmatrix} + \boldsymbol{I} & \boldsymbol{0} \\ \boldsymbol{0} & -\boldsymbol{I} \end{bmatrix}
    \begin{bmatrix} x_j \\ y_j \end{bmatrix}.
\end{equation}
Therefore, for any $i,j=1,\dots,k$, we must have
\begin{equation}
    \begin{bmatrix} x_i \\ y_i \end{bmatrix}^\dagger
    \begin{bmatrix} + \boldsymbol{I} & \boldsymbol{0} \\ \boldsymbol{0} & -\boldsymbol{I} \end{bmatrix}
    \begin{bmatrix} x_j \\ y_j \end{bmatrix} = \delta_{i,j},
\end{equation}
which means that eigenvectors generated from Eq.~\eqref{Eq:QRPAeigenvectorsWithARPACK}
are indeed the columns $X\boldsymbol{e}_i,Y\boldsymbol{e}_i$ of the $X,Y$ matrices.
In cases of degeneracies, the generalized \nicolas{Gram–Schmidt} orthogonalization procedure 
with QRPA metric is needed in order to orthogonalize the eigenvectors $\{x_i,y_i\}$ 
which belong to the degenerate eigenfrequency. It is important to emphasize that the 
numerical accuracy of the FAM solver, namely the mapping \eqref{Eq:QRPAmapping}, must 
be high enough to avoid loss of accuracy when computing the normalization factor 
$\sqrt{||u_i||^2-||v_i||^2}$ in those cases where the eigenfrequency is very close to 0.

Let us also point out the importance of carefully choosing the initial vector for 
the iterative algorithm in libraries such as ARPACK. This comes from the fact that 
typical FAM solvers assume certain symmetries such as, e.g. axialy symmetry, which 
means that only the excitation operators with certain selection rules are allowed. 
Typically this involves operators with well defined angular momentum $K$ and parity 
$\pi$, which actually means that one works in a subspace of the QRPA matrix with well 
defined $K^\pi$. Therefore we also need to supply the eigensolver with an initial 
random vector for the mapping \eqref{Eq:QRPAinverseMapping} that also satisfies the 
same selection rules as the ones of the FAM solver. In other words, the initial 
random vector must be an element of the appropriate $K^\pi$ subspace.

In order to illustrate the method described above on a realistic example, we 
select the axially deformed ($\beta_2\approx 0.5$) isotope ${}^{20}\mathrm{Ne}$, 
DD-PC1 Lagrangian and use a basis of 12 oscillator shells; see 
Ref.~\cite{mercier2021lowenergy} for a detailed discussion of the multipole 
response of this isotope. We implemented the spurious mode removal method in the 
\texttt{DIRQFAM} code \cite{bjelcic2020implementation,bjelcic2023implementation} 
using the ARPACK eigensolver, and focus on the isoscalar $J=2,K=1$ response which 
exhibits a rotational $K^\pi = 1^+$ spurious mode.

\begin{figure}[!htb]
\centering
\includegraphics[scale=0.35]{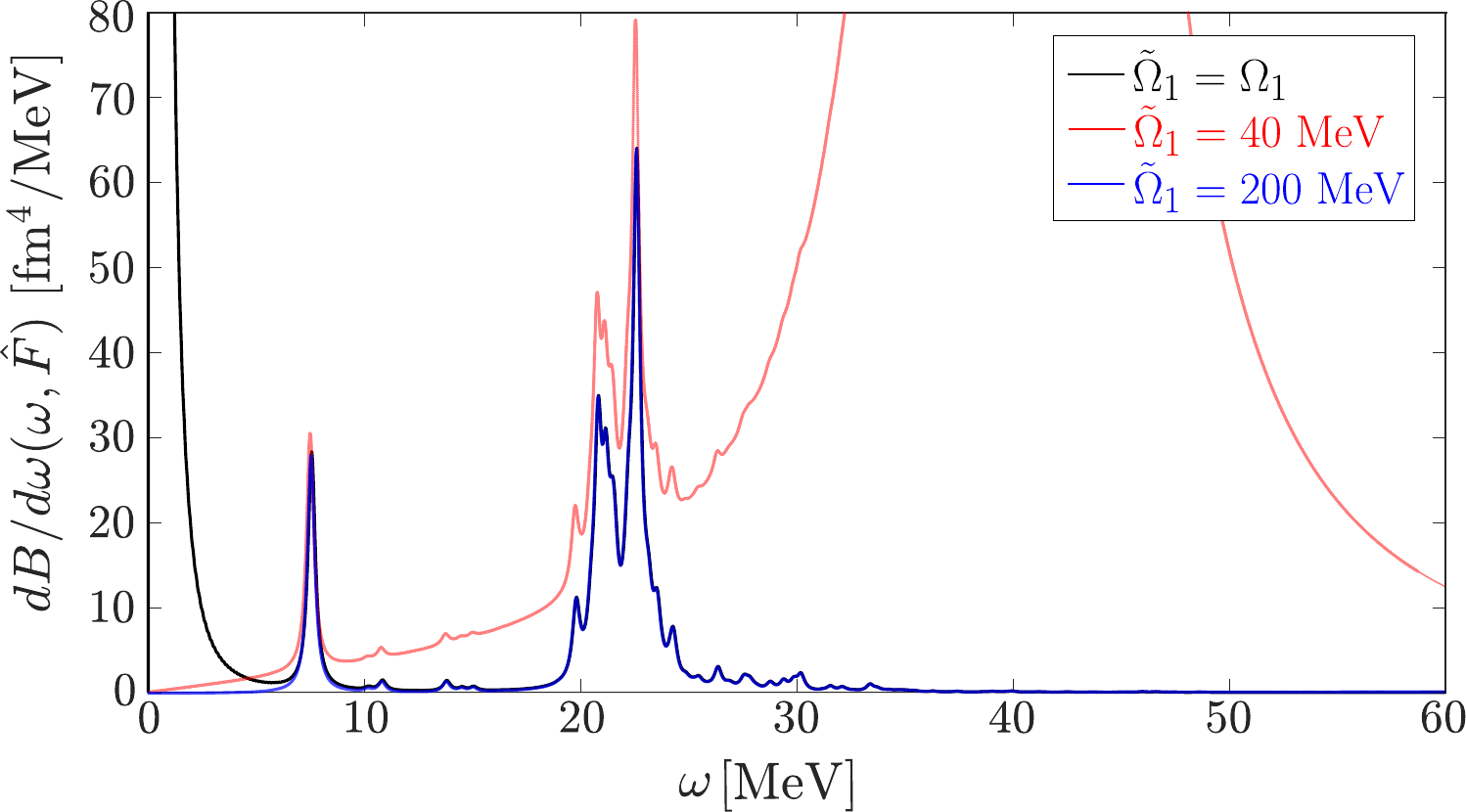}
\caption{Response function of ${}^{20}\mathrm{Ne}$ for isoscalar quadrupole 
$J=2,K=1$ excitation. The black curve shows the response without spurious mode 
removal, the red curve shows the response when the spurious mode is shifted to 
$\widetilde{\Omega}_1=40$ MeV, and the blue curve shows the response when the 
spurious mode is shifted to $\widetilde{\Omega}_1=200$ MeV.} 
\label{Fig:SpuriousModeRemoval}
\end{figure}

We generated a random $K^\pi = 1^+$ excitation operator as the initial residual 
vector by ARPACK. This ensures that the Arnoldi algorithm implemented in ARPACK 
is contained within the $K^\pi = 1^+$ subspace of the QRPA matrix. We computed 
the response function with the KPM method and bounding frequency $\Omega_b = 4500$ 
MeV, a Lorentz kernel with damping parameter $\lambda = 9$ and $N_{\mathrm{it}} = 
100000$ FAM iterations. This roughly corresponds to an equivalent smearing equal 
to $\gamma = \Omega_b \frac{\lambda}{2N_{\mathrm{it}}+1} \approx 0.2$ MeV; see 
Ref.~\cite{bjelcic2022chebyshev} for details.
In Fig.~\ref{Fig:SpuriousModeRemoval} 
we show the computed response functions:
the black curve shows the reponse without shifting the spurious mode, i.e.
$\widetilde{\Omega}_1 = \Omega_1$,
the red curve shows the response  when the 
spurious mode is shifted to $\widetilde{\Omega}_1=40$ MeV,
and the blue curve 
shows the response when the spurious mode is shifted to high-energy region $\widetilde{\Omega}_1 = 200$ MeV. Note that even though we could shift the energy of the spurious method 
to any arbitrary value $\widetilde{\Omega}_i$, it should still satisfy 
$\widetilde{\Omega}_i < \Omega_b$ for the KPM method to work.

\bibliographystyle{elsarticle-num}
\bibliography{zotero_output,books,others}

\end{document}